\documentclass{aa} 
\usepackage{color}
\usepackage{graphicx}
\usepackage{amsmath}
\usepackage{txfonts}
\usepackage{natbib}
\usepackage{url}
\usepackage{xspace}
\bibpunct{(}{)}{;}{a}{}{,}
\usepackage{lscape}
\usepackage{booktabs}
\usepackage[colorlinks=true,citecolor=blue,linkcolor=magenta,urlcolor=blue]{hyperref}
\newcommand{\teff}{$T_{\rm eff}$}

\newcommand{\cmap}{chromosome map}
\newcommand{\cmaps}{chromosome maps}
\newcommand{\cmag}{$C_{\rm F275W, F336W, F438W}$}
\newcommand{\met}{[M/H]}
\newcommand{\ngc} {NGC\,2808}
 \usepackage{lineno}
\usepackage{ulem} 

\begin{document}


\title{A stellar census in globular clusters with MUSE: Multiple populations chemistry in NGC\,2808\thanks{Based on observations collected at the European Organisation for Astronomical Research in the Southern Hemisphere, Chile (proposal IDs 094.D-0142(B), 096.D-0175(A))}
}

\author{M. Latour\inst{1}, T.-O. Husser\inst{1}, B. Giesers\inst{1}, S. Kamann\inst{2}, F. G\"ottgens\inst{1}, S. Dreizler\inst{1}, J. Brinchmann\inst{3,4}, N. Bastian\inst{2},
M.~Wendt\inst{5}, P. M. Weilbacher\inst{6}, and N. S. Molinski\inst{7,1} 
} 

\institute{Institut f\"ur Astrophysik, Georg-August-Universit\"at G\"ottingen, Friedrich-Hund-Platz 1, 37077 G\"ottingen, Germany\\ \email{marilyn.latour@uni-goettingen.de}
\and
Astrophysics Research Institute, Liverpool John Moores University, 146 Brownlow Hill, Liverpool L3 5RF, United Kingdom 
\and
Instituto de Astrof{\'\i}sica e Ci{\^e}ncias do Espaço, Universidade do Porto, CAUP, Rua das Estrelas, PT4150-762 Porto, Portugal
\and
Leiden Observatory, Leiden University, P.O. Box 9513, 2300 RA, Leiden, The Netherlands
\and
Institut f\"ur Physik und Astronomie, Universit\"at
Potsdam, Karl-Liebknecht-Str. 24/25, 14476 Golm, Germany
\and
Leibniz-Institut f\"ur Astrophysik Potsdam (AIP), An der Sternwarte 16, 14482 Potsdam, Germany
\and
Institut f\"ur Geophysik und extraterrestrische Physik, Technische Universit\"at zu Braunschweig, Mendelssohnstr. 3, 38106 Braunschweig, Germany 
}

\date{Received ; accepted}

\abstract
{Galactic globular clusters (GCs) are now known to host multiple populations displaying particular abundance variations. The different populations within a GC can be well distinguished following their position in the pseudo two-colors diagrams, also referred to as "chromosome maps." These maps are constructed using optical and near-UV photometry available from the \textit{Hubble} \textit{Space} \textit{Telescope} (HST) UV survey of GCs. However, the chemical tagging of the various populations in the chromosome maps is hampered by the fact that HST photometry and elemental abundances are both only available for a limited number of stars. }
   {The spectra collected as part of the MUSE survey of globular clusters provide a spectroscopic counterpart to the HST photometric catalogs covering the central regions of GCs. In this paper, we use the MUSE spectra of 1115 red giant branch (RGB) stars in NGC\,2808 to characterize the abundance variations seen in the multiple populations of this cluster.   }
   {We used the chromosome map of NGC\,2808 to divide the RGB stars into their respective populations. We then combined the spectra of all stars belonging to a given population, resulting in one high signal-to-noise ratio spectrum representative of each population.}
   {Variations in the spectral lines of O, Na, Mg, and Al are clearly detected among four of the populations. In order to quantify these variations, we measured equivalent width differences and created synthetic populations spectra that were used to determine abundance variations with respect to the primordial population of the cluster. Our results are in good agreement with the values expected from previous studies based on high-resolution spectroscopy. We do not see any significant variations in the spectral lines of Ca, K, and Ba. We also do not detect abundance variations among the stars belonging to the primordial population of NGC\,2808. 
   } 
   {We demonstrate that in spite of their low resolution, the MUSE spectra can be used to investigate abundance variations in the context of multiple populations. }
   
 \keywords{globular clusters: individual: NGC\,2808 --- Stars: abundances --- Techniques: imaging spectroscopy }  
   
 \authorrunning{Latour et al.}
\titlerunning{Multiple populations chemistry in NGC 2808}

\maketitle

\section{Introduction}

Galactic globular clusters (GCs) have been traditionally viewed, and modeled, as simple stellar populations made of stars sharing the same evolutionary history. However, some particular properties of GC stars indicate that the story is not that simple. For example, it has become clear that nearly all Galactic GCs host significant abundance spreads among their stars and some patterns are ubiquitous among GCs, such as the \mbox{Na-O} and \mbox{N-C} anticorrelations (see reviews by Gratton et al. \citeyear{gra04,gra12}). These abundance anomalies are characteristics of globular clusters and are not observed in large numbers ($\sim$3\% in the halo, $\sim$7\% in the bulge) among stars of the Galactic field \citep{martell10,carr10,schia17,koch19}.  
Additional evidence pointing to a more complex stellar formation scenario came with the observation of bimodal regions in the color-magnitude diagram (CMD) of GCs.
Such bimodal distributions have been found along the main sequences (MSs), subgiant branches (SGBs), and red giant branches (RGBs) of several clusters and are mainly caused by variations in the strength of molecular bands, such as CN and NH, that affect the stellar flux in the UV and blue optical regions (see, e.g., \citealt{piotto07,mil08,han09,piotto12}).
Changes in these spectral features can be detected, when using the appropriate filters, with \textit{Hubble} \textit{Space} \textit{Telescope} (HST) and ground-based wide band photometry \citep{monelli13,mass16,nieder16}. Targeted photometry with narrow and/or middle band filters like the Washington system \citep{cumm14}, the Ca-CN system \citep{lee19}, and specific narrow-band HST filters \citep{larsen14} also allow the detection of multiple sequences in CMDs.

These complex structures seen with photometry and the abundance variations measured spectroscopically are in fact related. It has been shown that the color differences between the multiple RGBs, SGBs, and MSs are related to differences in the abundances of some specific elements (mostly He, C, N, and O), and in fewer cases by differences in iron abundances (see, e.g., \citealt{mar12,mil13_6752,yong15,bell17III,lar18,mil18}). With such compelling evidence, it is now accepted that nearly all GCs older than about 2 Gyrs \citep{mart18,bas18} host multiple populations that can be distinguished by their different photometric and/or spectroscopic properties. 
Although by now many observational studies have measured abundances and characterized the relation between the variations of different elements (e.g., He, Li, C, N, O, Na, Mg, and Al) in the stars of many globular clusters, the enrichment mechanism(s) responsible for such particular abundance variation patterns is still hotly debated. With growing observational constraints to reproduce, the various enrichment mechanisms and stellar polluters proposed are currently unable to fulfill all the requirements (see \citealt{bas18} for a recent review on the topic).

The pseudo-two-color diagrams introduced by Milone et al. (\citeyear{mil15_2808}) and then termed as chromosome maps \citep{mil16} have proven to be a robust way to distinguish the various populations of a given GC, especially for stars on the RGB. 
These maps are built using a combination of HST filters ($F275W$, $F336W$, $F438W$, and $F814W$) that are sensitive to spectral features affected by the chemical variations characterizing the different populations.
\citet{mil17} presented the \cmaps\ of the 57 clusters included in their HST UV Legacy Survey of Galactic GCs \citep[HUGS;][]{piotto15} and showed that, for the majority of their clusters, the RGB stars can be easily divided into two main groups, which they refer to as first (1G) and second (2G) generations. Indeed, some overlap between stars in the \citet{mil17} sample and previous spectroscopic studies indicate that stars belonging to the 1G have a primordial chemical composition, while the abundances of the 2G stars show traces of processed material, for example Na enrichment and O depletion \citep{mil15_2808,omall18,cab19}. More recently, \citet{marino19} retrieved  spectroscopic abundances from literature studies for stars in the \cmaps\ of 29 GCs, confirming that stars belonging to the primordial population (or 1G) have light-element abundances similar to those of field stars, while the 2G stars are enhanced in N, Na, and depleted in O. 
However, the overlap between stars that have the optimal photometric data required to separate the populations and those that have spectroscopic abundances is limited (often less than 20 stars per clusters) given that the HST survey covers the central regions of the clusters, while spectroscopic surveys often target stars in the outskirt regions in order to avoid crowding issues. 

As part of the guaranteed time observations (GTO) with the integral-field spectrograph MUSE \citep{bacon10}, our team is carrying out a survey of Galactic globular clusters, especially targeting the clusters central regions. 
The overlap between our data and the HST photometry is ideal to associate stars with their respective populations according to their position in the \cmaps, but the low resolution of the MUSE spectra is not optimal to derive abundances for individual stars. Instead, we followed a different approach that consists in combining the spectra of the stars belonging to a given population. 
In this paper, we present and test our approach with the globular cluster \ngc. 
It is one of the few clusters to have an elaborate set of populations, based on its \cmap\ and on the abundance pattern of its RGB stars, but no significant spread in metallicity (or [Fe/H]). 
Because of its richness and complexity, the multiple populations of NGG\,2808 have been thoroughly studied in the past based on their photometric properties (e.g., \citealt{mil15_2808,lar18}) and their chemical abundances (e.g., Carretta et al. \citeyear{carr06,carr15_2808}, and \citealt{cab19}). 
In Sect. 2, we present our observational material, consisting of the MUSE spectroscopic sample and the HST photometric catalog. The methods used to derive atmospheric parameters, combine spectra and estimate abundance variations are explained in Sect. 3. 
Our results are presented in Sect. 4, where we compare them with expectations from literature studies and explore further population divisions in the \cmap.  A short conclusion follows in Sect. 5.

\section{Observational material }

\subsection{Spectroscopy}

\begin{table}
 \caption{Summary of MUSE observations of \ngc }\label{obs}
 \centering
 \scriptsize
 \begin{tabular}{c c c c c}
 \toprule
 \toprule
 Pointing & RA & DEC & Obs. date & Seeing \\
  & & & (UT) & (\arcsec) \\
 \midrule
 1 & 09:11:59.574 & $-$64:52:11.13 & 2014-12-18 08:06:36 & 1.04 \\
  & & & 2014-12-19 07:37:20 & 0.80 \\
   & & & 2016-03-13 02:48:52 & 0.82 \\
 2 & 09:11:59.562 & $-$64:51:26.13 & 2014-12-18 08:09:55 & 1.10 \\
   & & & 2014-12-19 07:40:42 & 0.84 \\
   & & & 2016-03-13 03:01:24 & 0.90 \\
 3 & 09:12:06.639 & $-$64:52:11.06 &  2014-12-18 08:13:15 & 1.10 \\
    & & & 2014-12-19 07:44:05 & 0.82 \\
   & & & 2016-03-14 00:49:53 & 0.90 \\
 4 & 09:12:06.623 & $-$64:51:26.13 & 2014-12-18 08:16:35 & 1.32 \\
   & & & 2014-12-19 07:47:27 & 0.76 \\
   & & & 2016-03-14 01:01:56 & 0.84 \\
 \bottomrule                                
\end{tabular}
\end{table}

The observations of NGC\,2808 were performed as part of the MUSE GTO dedicated to globular clusters (PI: S. Dreizler, S. Kamann). 
So far, the spectroscopic data collected as part of this survey have been used for various purposes, such as kinematic analyses \citep{kam18}, characterizing binary systems \citep{gie18}, and the search for emission line objects \citep{goet19,goett_elo}. 
A detailed description of the program, as well as the data reduction and their analysis is provided in \citet{kam18}. In the following, we briefly summarize the information specifics to the observations of \ngc. 

\begin{figure}
 \includegraphics[width=\columnwidth]{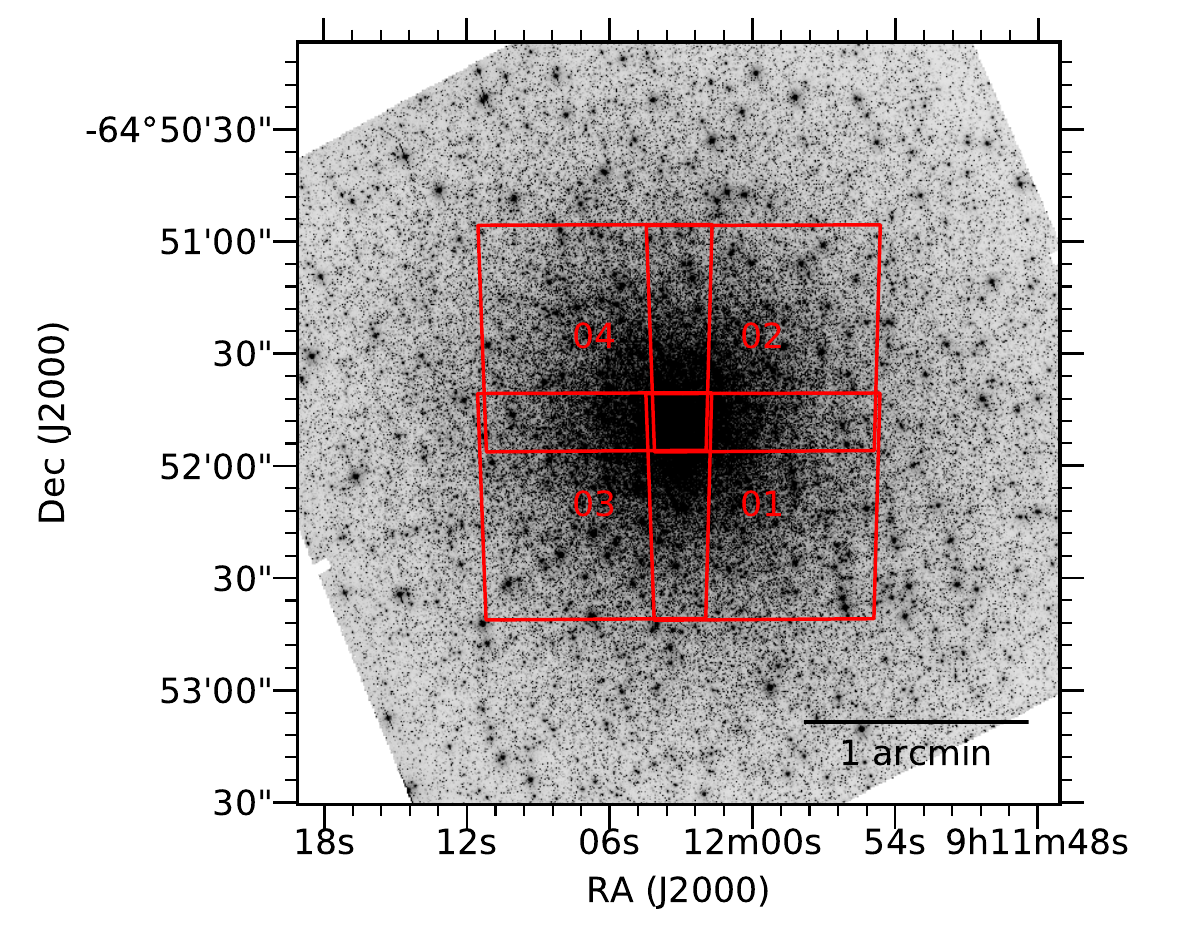}
 \caption{Position of the four pointings of MUSE observations. The background shows an HST/ACS F606 image of \ngc.}
 \label{pointings} 
\end{figure}

The central region of the cluster is covered by a mosaic consisting of the four pointings shown in Fig.~\ref{pointings}. The data were obtained with the wide field mode of MUSE, that provides a field of view of 1\arcmin $\times$ 1\arcmin. 
For this paper, we used the spectra collected until March 2016, thus preceding the commissioning of the adaptive optic system installed on UT4 of the VLT.
Each pointing was observed at three different epochs. The observations listed in Table~\ref{obs} consisted of three exposures, offset by 90 degrees in the derotator angle. 
The individual exposures were processed with the standard MUSE pipeline (\citeauthor{weil12}, \citeyear{weil12,weil14}) which was also used to create a combined data cube for each observation. The total integration time per observation is 495\,s. The spectra of the individual stars were extracted from the datacube using the \textsc{PampelMuse} software described in \citet{kam13}. 
The extraction of the spectra relies on the existence of a photometric catalog that includes astrometry and photometry of sources in the field of view. For \ngc, we used the HST data from the Advanced Camera for Surveys (ACS) of Galactic
globular clusters \citep{sarajedini07,and08}.
The resulting spectra cover the 4750$-$9350\,\AA\ wavelength range with an 
average spectral resolution of $\sim$2.5\,\AA, although this varies slightly across the wavelength range \citep{huss16}.

\begin{figure*}
 \includegraphics[width=\textwidth]{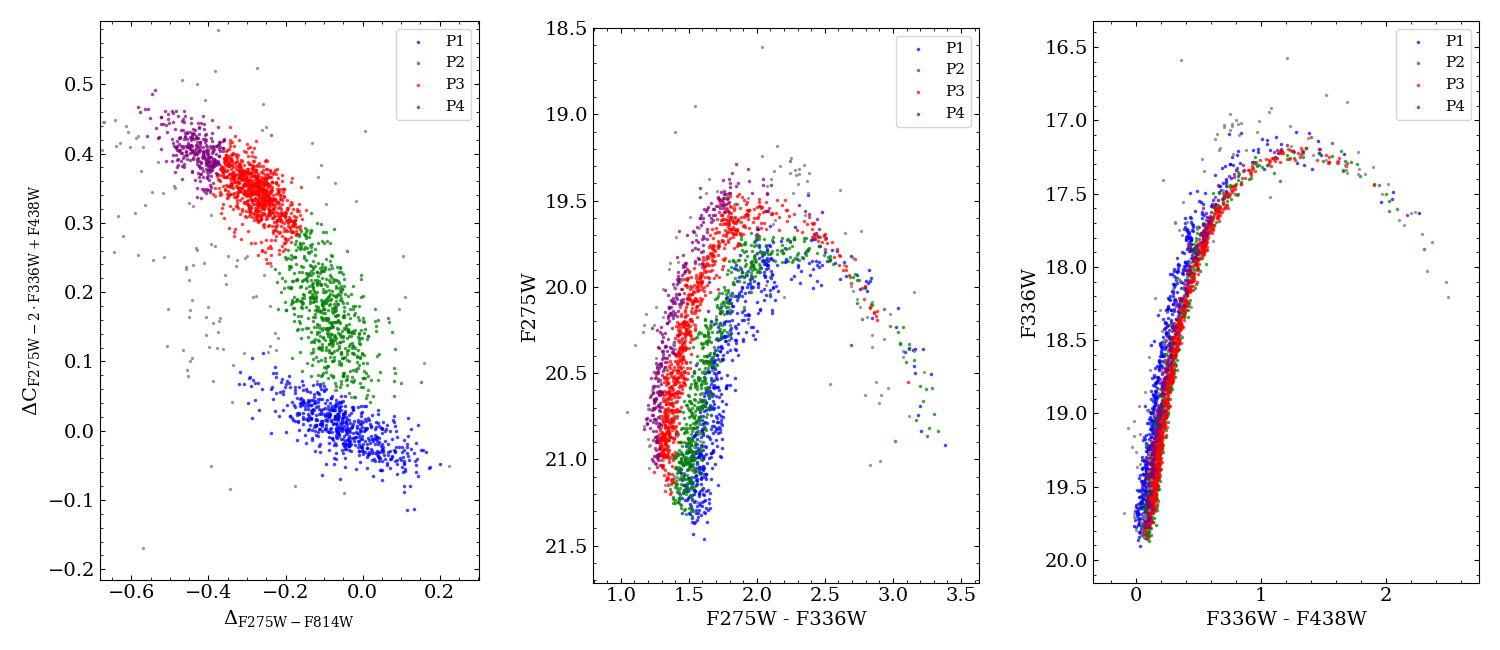}
 \caption{Chromosome map and CMDs of NGC\,2808 with its four populations. The stars plotted in gray are not included in a population.}
 \label{cmap_2808}
\end{figure*} 

\subsection{Photometry}

To create the \cmap, we used the photometric data published in the astrophotometric catalogs of the  HST UV Globular cluster Survey (HUGS) presented by \citet{nar18}. Three versions of the catalogs are available for each cluster, corresponding to three different methods of extracting the data (see \citealt{bell17I}). For our multiple populations study involving RGB stars, we used the catalogs corresponding to method\,1, which is optimal for bright stars. 

In a first step, we clean the HST photometry following the procedure described in \citet{nar18}. This cleaning procedure allows the selection of stars with well-measured photometry according to their photometric error, the shape of their PSF and the quality of their point spread function (PSF) fit during extraction. Our final photometric sample includes only stars that pass the selection criteria for these three parameters in all four filters involved in the construction of the chromosome maps. Although this decreases the number of stars left to work with in the subsequent steps, the resulting chromosome maps are cleaner and the different populations are less likely to be contaminated by stars with uncertain photometry. 
The \cmap\ was constructed following the method described in \citet{mil17}. We defined the RGB envelope along magnitude bins in the F814W filter in a way that the blue fiducial line is at the 10$^{\rm th}$ percentile and the red fiducial line at the 90$^{\rm th}$ percentile. The transformation from the CMD ($m_{\rm F814W}$, $m_{\rm F275W}-m_{\rm F814W}$) and pseudo-CMD ($m_{\rm F814W}$, \cmag) to the \cmap\ requires a value for the width of the RGB, which we compute as the mean difference between the red and blue fiducial lines defining the RGB envelope. 

Figure \ref{cmap_2808} shows our chromosome map of the cluster. Our chromosome map is very similar to the one presented in \citet{mil17} even if we did not correct the photometry for differential reddening.
Following the classification of \citet{mil17}, the stars of a globular cluster can be divided into two main groups in the \cmap. The population\,1 stars are found at the bottom of the \cmap, around the (0,0) position, while population\,2 stars extend above\footnote{We note that \citet{mil17} used the term generation instead of population and referred to them as 1G and 2G stars.}. However, a handful of clusters, such as \ngc, host a more complex population of RGB stars and our selection of four populations (P1 to P4) is based on those identified in \citet{mil15_2808}.
When defining the populations, we aim at selecting stars that can be clearly assigned to one of the populations, thus we leave out stars whose positions are scattered around in the \cmap. Along with the \cmap, we also plot the position of the stars in two different CMDs. The $m_{\rm F336W}-m_{\rm F438W}$ color provides a clear distinction between the population\,1 and 2 stars as defined by \citet{mil17}. Even though the population\,2 stars in \ngc\ hold distinct sub-populations, they are not discernible in this particular color. 
Although previous studies have divided our P1 in subgroups \citep{mil15_2808,lar18}, we consider these stars as a single population and will discuss the subdivisions within P1 in Sect. 4.3.

\section{Methods}

Studies on the abundances of RGB stars in globular clusters have mostly been performed using high resolution ($R$ $\sim$20~000 $-$ 40~000) spectra (see, e.g., \citealt{carr06} and following papers in that series). In order to reliably derive abundances of individual elements in these stars, it is necessary to resolve the lines of interest to avoid uncertainties due to blending. With their low resolution, the MUSE spectra are not well-suited for such an analysis on individual stars. However the lack in resolution can be compensated, to some extent, by the large amount of stars we have in our sample. After matching our MUSE sample with the stars in the chromosome map of \ngc, we obtained a sample of more than 1100 RGB stars with an assigned population. Our approach is thus to combine the spectra of all stars in a given population and use the resulting high signal-to-noise (S/N) spectrum to represent the whole population. We then searched for abundance variations by comparing the spectra of the different populations. Assessing chemical abundances from our populations spectra, however, is not a straightforward task. Measuring absolute abundances would be hampered by the fact that many lines are strongly blended, either with other photospheric lines or with interstellar absorption (e.g., the sodium D doublet). Instead we attempt to estimate differential abundances between the populations using the ``primordial" population (P1) as a reference. We present a description of the different steps required to achieve this goal in the following subsections. 

\subsection{Atmospheric parameters determination}
Atmospheric parameters are obtained for all individual spectra following a procedure similar to that described in \citet{huss16}. Firstly, we find an isochrone \citep[from][]{marigo17} that best matches the HST photometry (F606W, F606W-F814W) from \cite{sarajedini07}. For \ngc, the best-matching isochrone has an age of 12 Gyr and [M/H] = -0.93. Secondly, we derive $T_\mathrm{eff}$ and $\log g$ values for all stars by finding the nearest point on the isochrone in the CMD. 
These values (and the mean metallicity of the cluster) are then used to get a template spectrum from the G\"ottingen spectral library of PHOENIX spectra \citep{huss13} and perform a cross-correlation. Finally, the atmospheric parameters from the isochrone and the radial velocity from the cross-correlation are used as initial values to run a full-spectrum fit against the full grid of PHOENIX spectra using the spectrum fitting framework \textit{spexxy}\footnote{https://github.com/thusser/spexxy}. The best fit provides $T_\mathrm{eff}$, [M/H], $v_\mathrm{rad}$, and a model for the telluric lines for all observed spectra. The surface gravity is kept fixed to the value from the isochrone during this process because $\log g$ cannot be well constrained using low-resolution spectra.

All stars in our sample have up to three observed spectra, so we combine the results from the fits of the individual spectra to get final parameters for every star. During this step, we evaluate the reliability of each spectrum following the method described in Section 3.2 of \citet{giesers19}. This method evaluates the quality of the observed spectra based on S/N, extraction results, and radial velocity measurements. 
The main difference with \citet{giesers19} is that we require the reliability (R$_{\rm total}$ value) on the radial velocity to be at least 50\% (instead of the more restrictive 80\% required for the binary studies).  
In the end, the parameters ($T_\mathrm{eff}$ and [M/H]) obtained from the individual spectra of a given star satisfying the reliability criteria are used to compute the weighted average values that are adopted as $T_\mathrm{eff}$ and [M/H] of the star.

\subsection{Spectral combination}
In order to get a good, high S/N spectrum for every star, we combine all its observed spectra. At first, we remove the telluric lines from the raw spectra by dividing them by the telluric model obtained in the full-spectrum fit. Because the extracted spectra from the MUSE cubes are not perfectly flux calibrated, the full-spectrum fit also produces a polynomial that describes the difference between the observed spectrum and the model (i.e., similar to a continuum if the models were normalized). We also divide each spectrum by this polynomial to get rid of the uneven continuum (see, e.g., Fig. 16 of \citealt{huss16}). Then we shift all spectra to rest-frame using the obtained radial velocity and resample them to the same wavelength grid. Finally, we co-add the individual spectra of each star, using their S/N value as weight.

We create the populations spectra by adding the fluxes of each star. Because the exposure time is similar for all stars and the observed spectra are flux calibrated, the brighter stars have a higher flux and a higher S/N than the fainter stars. Therefore the direct summation of the fluxes ensures that the lower S/N spectra contribute less to the final population spectrum and provides a natural S/N weighting.  
During the combination process, we reject stars for which our membership probability, based on metallicity and radial velocity (see \citealt{kam18}), is below 80\% \footnote{We note that only one star was rejected as a non-member.} and also stars that are identified as emission line objects \citep{goett_elo}. We also excluded from our sample stars whose spectrum has a S/N < 20. 
Because we intend to model the RGB stars for abundance variations, we also excluded the most luminous stars (log $g$ < 1.0 and \teff\ < 4500 K) at the tip of the RGB where sphericity, wind, and non-LTE effects are expected to be important and our synthetic spectra might not be appropriate. After applying this selection, our sample consists of 1115 RGB stars included in the four populations.

\subsection{Computation of synthetic spectra}

We computed synthetic spectra with varying elemental abundances using the latest version of the SYNSPEC code (version 53, I. Hubeny, priv. comm.; \citealt{hub11,hub17}) in combination with the atmospheric structures of the PHOENIX models from the G\"ottingen spectral library. SYNSPEC is a general spectrum synthesis program that solves the radiative transfer equation for a given atmospheric structure. It was originally developed to be used in conjunction with TLUSTY, a non-LTE stellar atmosphere code \citep{hub95}, but it can also readily use an input model in the Kurucz's \textsc{atlas} format, a feature we made use of by converting the PHOENIX models into the \textsc{atlas} format. The latest version of SYNSPEC has been upgraded to provide a better treatment of molecular opacities that are important for the computation of cool stars spectra (Hubeny et al., in preparation). We used the atomic and molecular line lists provided with the TLUSTY and SYNSPEC codes that are based on the Kurucz data. By comparing some of our synthetic spectra with the PHOENIX spectra, we realized that the updated atomic data for \ion{Fe}{I} \citep{K14} and \ion{Nd}{I} \citep{HLSC} retrieved from the VALD database \citep{vald15} provided a better match and we updated the line lists accordingly. For the lines of interest in the context of abundance variations (see Table \ref{llist}), we verified, and updated if necessary, their atomic data following the VALD and NIST \citep{NIST_ASD} databases.

For every atomic element inspected, we computed synthetic spectra with varying abundances of the given element in steps of 0.25 dex.
This was done using the PHOENIX model atmosphere at [M/H] = $-$1.0, which is very close to the average [M/H] that we obtained for the RGB stars in \ngc\ ($-$1.03 dex, see Sect. 4.1). 
As for the models used to derive the atmospheric parameters, we assumed a scaled-solar metallicity for the abundance of all other elements. 
A set of synthetic spectra (with varying abundances) was interpolated for every star at its \teff\ and log $g$ value. In order to combine the synthetic spectra using appropriate weights, we multiplied the normalized model spectra of each star by the average flux of its MUSE spectrum.
The procedure results in one set of spectra with varying abundance for each population. 

\subsection{Equivalent widths measurements}

We measured equivalent width (EW) differences of spectral lines
by integrating over the residuals obtained when subtracting the spectrum of P1 from that of the other populations (see Sect. 4.2). 
By working with EW differences, we eliminate the contributions of blended features whose strength can be assumed to be constant between the populations (e.g., Fe lines). We also eliminate the contribution of the interstellar medium (ISM) component of the  sodium D (NaD) lines,  assuming the stars of each population are equally distributed in the field of view. Although
\cite{wendt17} found that the 
strength of the NaD (and \ion{K}{i}) ISM varies across the field of view in NGC\,6397, the preliminary results for \ngc\ do not show a strong spatial variation (Wendt et al., in preparation). 
We computed the EWs using a trapezoidal integral because most of the residuals are too coarsely sampled to fit them with a line profile. The errors on the EWs are estimated by doing a similar exercise over spectral regions of constant strength between the populations. We selected 12 such reference regions and used the average of their EW differences (in absolute value) as uncertainty, resulting in uncertainties between 8$-$13\,m\AA\ depending on the population.

\section{Results}

\subsection{Stellar parameter distributions}

\begin{figure}
\includegraphics[width=\columnwidth]{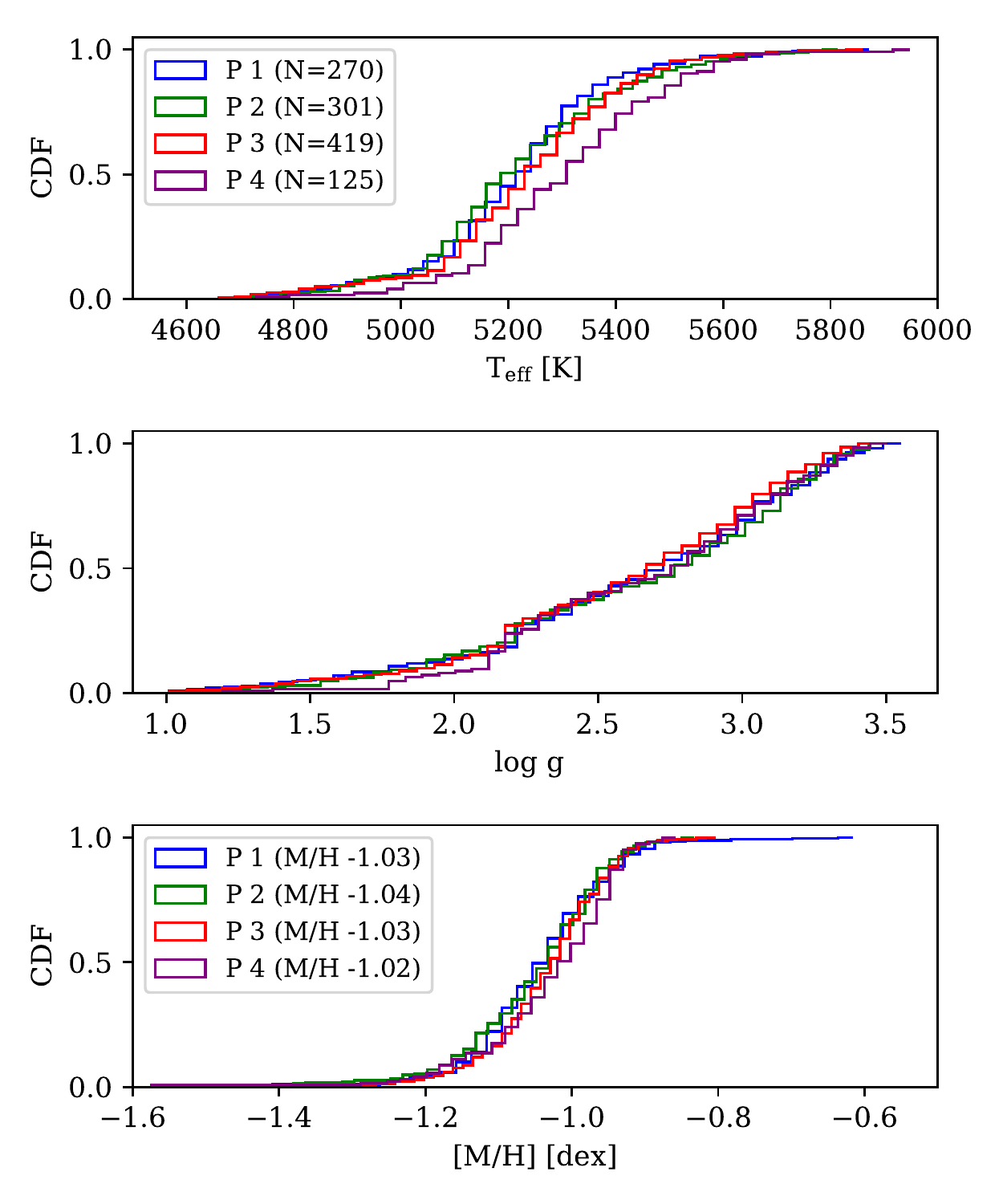}
 \caption{Cumulative distributions in \teff, log $g$, and \met\ of the stars included in the four populations of NGC\,2808. The number of stars included in each population is indicated in the legend of the upper panel while the average \met\ value for each population is indicated in the legend of the lower panel.}
 \label{dist_2808}
\end{figure}

\begin{table}[ht!]
 \caption{List of atomic lines of interest covered by the MUSE wavelength range}\label{llist}
 \centering
 \small
 \begin{tabular}{l c c }
 \toprule
 \toprule
 Element & Wavelength & Note \\
 & (\AA) & \\
 \midrule
 \ion{O}{i}*	& 7774.17	&  \\

 \ion{Na}{i}	& 5682.63	&  B \\
 \ion{Na}{i}	& 5688.19	&  B \\
 \ion{Na}{i}	& 5688.21	&  B \\
 \ion{Na}{i}/(NaD)*	& 5889.95 	&  R, B with ISM\\
 \ion{Na}{i}/(NaD)*	& 5895.92	&  R, B with ISM \\
 \ion{Na}{i}	& 6154.22	&  B \\
 \ion{Na}{i}	& 6160.74	&  B \\
 \ion{Na}{i}	& 8183.25	&  \\
 \ion{Na}{i}	& 8194.80	&  \\
 
 \ion{Mg}{i}/(Mg $b$)*	& 5167.32	& B \\
 \ion{Mg}{i}/(Mg $b$)*	& 5172.68	& B \\
 \ion{Mg}{i}/(Mg $b$)*	& 5183.60	& B \\
 \ion{Mg}{i}	& 5711.08	& B \\
 \ion{Mg}{i}	& 7657.60	& B \\
 \ion{Mg}{i}	& 7659.12	& B \\
 \ion{Mg}{i}	& 7691.55	&  \\ 
 \ion{Mg}{i}	& 8736.00	&  \\ 
 \ion{Mg}{i}*	& 8806.75	&  \\ 

 \ion{Al}{i}*	& 6696.01	&  \\
 \ion{Al}{i}*	& 6698.67	&  \\
 \ion{Al}{i}	& 6906.40	&  \\
 \ion{Al}{i}	& 7083.96 	& B \\
 \ion{Al}{i}	& 7084.64 	& B \\
 \ion{Al}{i}*	& 7361.56	& B \\
 \ion{Al}{i}*	& 7835.31	& B \\
 \ion{Al}{i}*	& 7836.13	& B \\
 \ion{Al}{i}*	& 8773.88	& B \\
 
 \ion{Si}{i}	& 8648.46	& \\
 \ion{Si}{i}	& 8752.00	&  \\
 
 \ion{K}{i}	& 7664.90	& R, B with ISM \\
 \ion{K}{i}	& 7698.96	& R, B with ISM \\
 
 \ion{Ca}{i}	& 6161.30	& B  \\
 \ion{Ca}{i}	& 6162.17	& B \\
 \ion{Ca}{ii}/(CaT)	& 8498.02	&  \\ 
 \ion{Ca}{ii}/(CaT)	& 8542.09	&  \\
 \ion{Ca}{ii}/(CaT)	& 8662.14	&  \\
 
 \ion{Ba}{ii}	& 4934.08	& R, B \\ 
 \ion{Ba}{ii}	& 6141.71	& B \\
 \ion{Ba}{ii}	& 5853.67 	& B \\
 \ion{Ba}{ii}	& 6496.89	& B \\
 \bottomrule                                
\end{tabular}
\tablefoot{Note. Transitions marked with an * were used to derive abundance differences. B - Blended with other strong lines (at the MUSE resolution), 
R - Resonance lines.}
\end{table}

\begin{figure*}
 \includegraphics[width=\columnwidth]{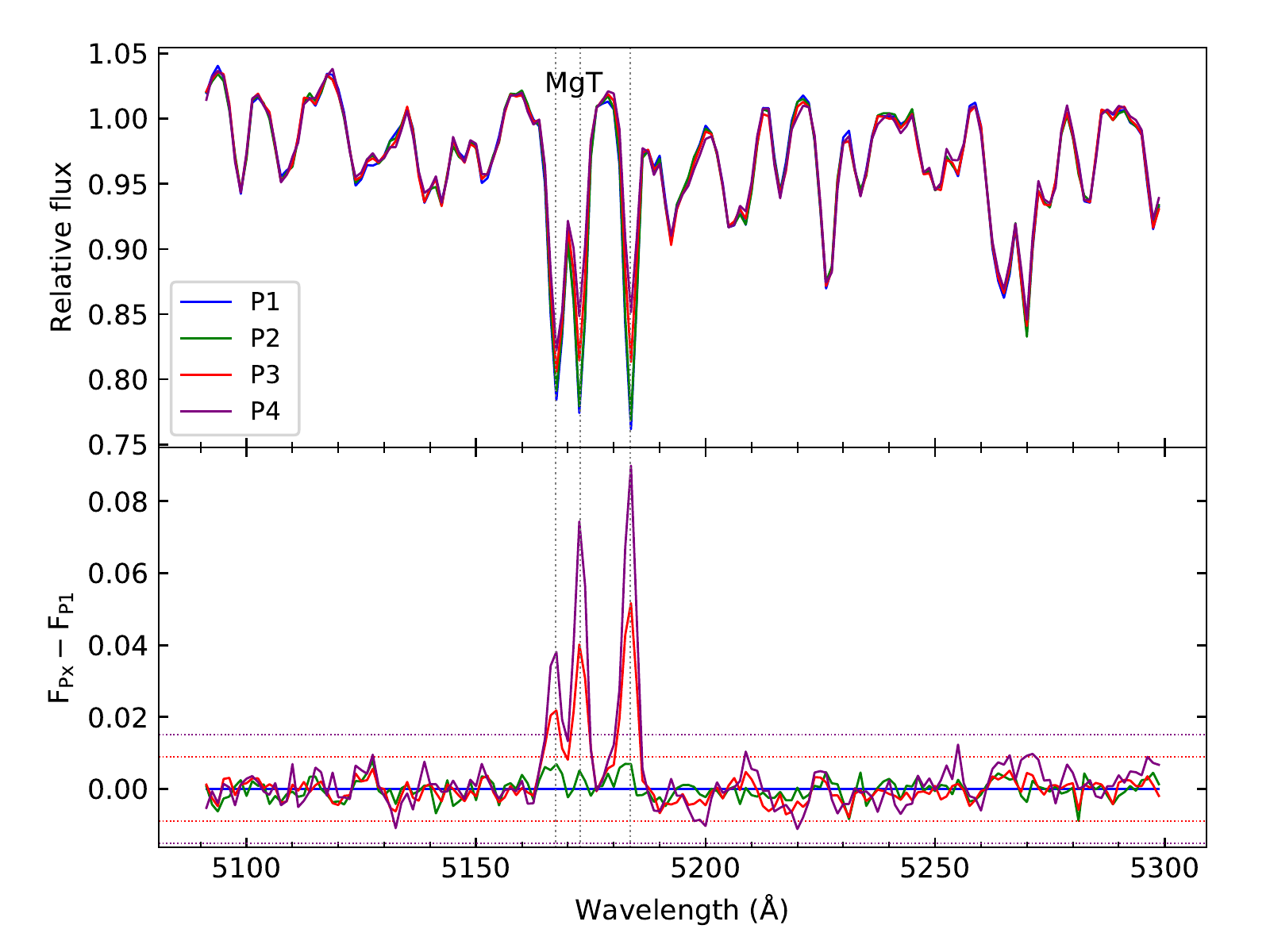}
 \includegraphics[width=\columnwidth]{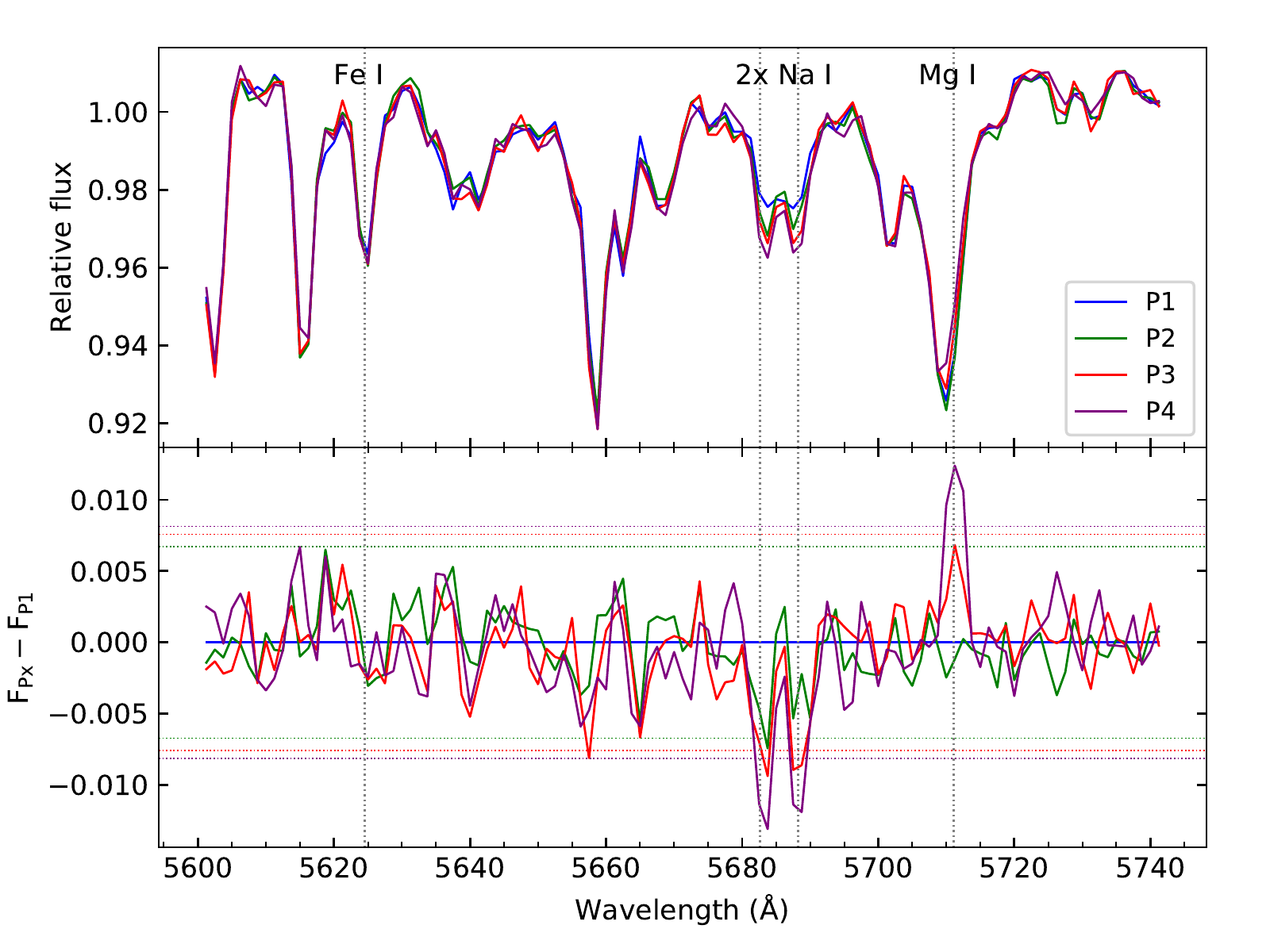}
 \includegraphics[width=\columnwidth]{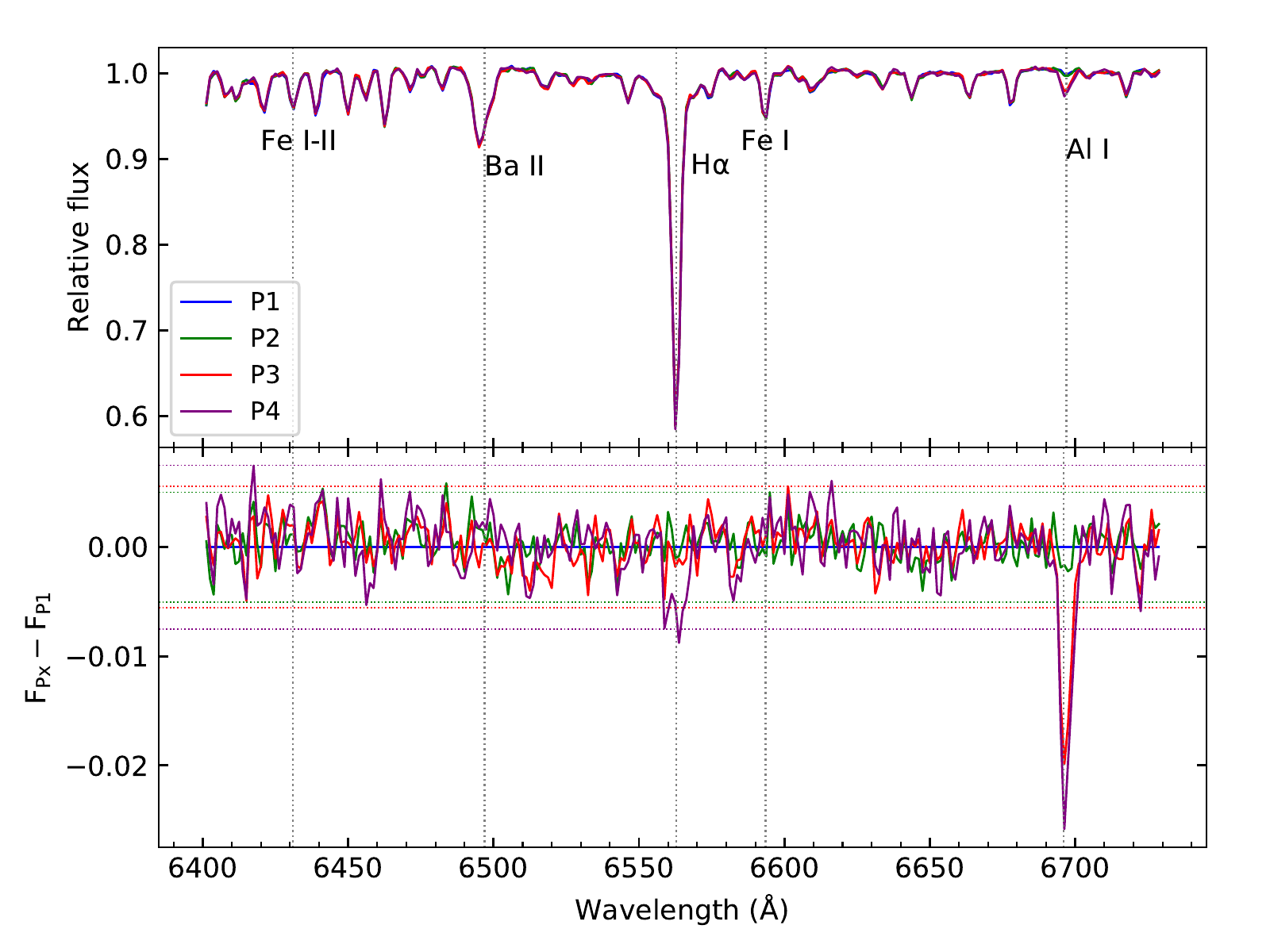}
 \includegraphics[width=\columnwidth]{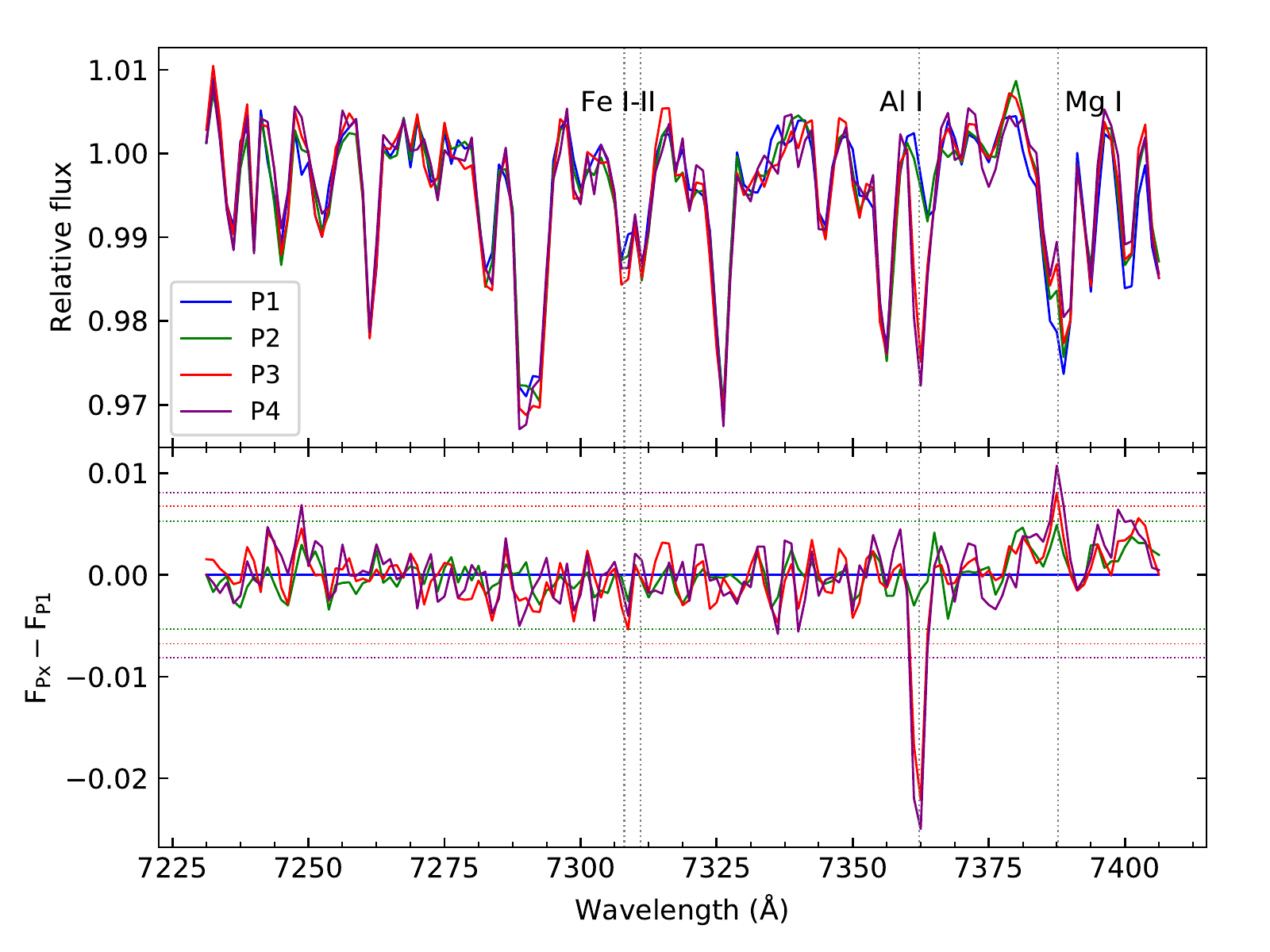} .
   \caption{Comparisons between the spectra of the four populations in NGC\,2808. The residuals on the bottom panels are plotted as the difference between the flux of a given population and that of P1 (F$_{Px}$ - F$_{P1}$). The horizontal dotted lines represent the 3$\sigma$ value of the residuals of each population over the plotted range. }
  \label{spec_2808} 
  \end{figure*}
  
Figure \ref{dist_2808} shows the cumulative distribution functions in \teff, log $g$, and \met\ of the stars included in each of the four populations as obtained from the fitting procedure described in Sect 3.1.  
The parameter distributions of the populations are overall very similar. One conspicuous difference is seen in the \teff\ distribution of P4 that appears to contain hotter stars on average. This population has been considered as the most He-enhanced population in \ngc\ \citep{mil15_2808} and stars having different helium content also have different \teff\ and log $g$ at a given luminosity \citep{sbor11}. In fact He-enhanced stars are expected to be hotter at a given luminosity and \citet{mil15_2808} estimated a \teff\ difference of about 100 K between the most He-enriched population (equivalent to our P4) and the population having a primordial helium content (equivalent to our P1). As for the difference in surface gravity, they estimated a more marginal change of about 0.05 dex. The difference in \teff\ that we observe between P1 and P4 (95 K at the median value of the distribution) is comparable to the expected effect and could be an indirect signature of the He-enhancement. However, as seen in the log $g$ distribution, P4 is lacking stars at the luminous (and thus cold) end of the RGB. Recomputing the \teff\ difference between P1 and P4 including only stars with log $g$ > 2 resulted in a smaller value of 70 K. 
A few clusters are known to have populations with different [Fe/H] and this would be seen in the metallicity distributions (Husser et al. 2019, submitted to A\&A). However, the stars in NGC\,2808 are expected to have the same iron content \citep{carr06} and that is reflected in the similar average metallicity that we obtained for each population.

\subsection{Abundance variations among the populations}

   \begin{figure*}
  \includegraphics[width=\columnwidth]{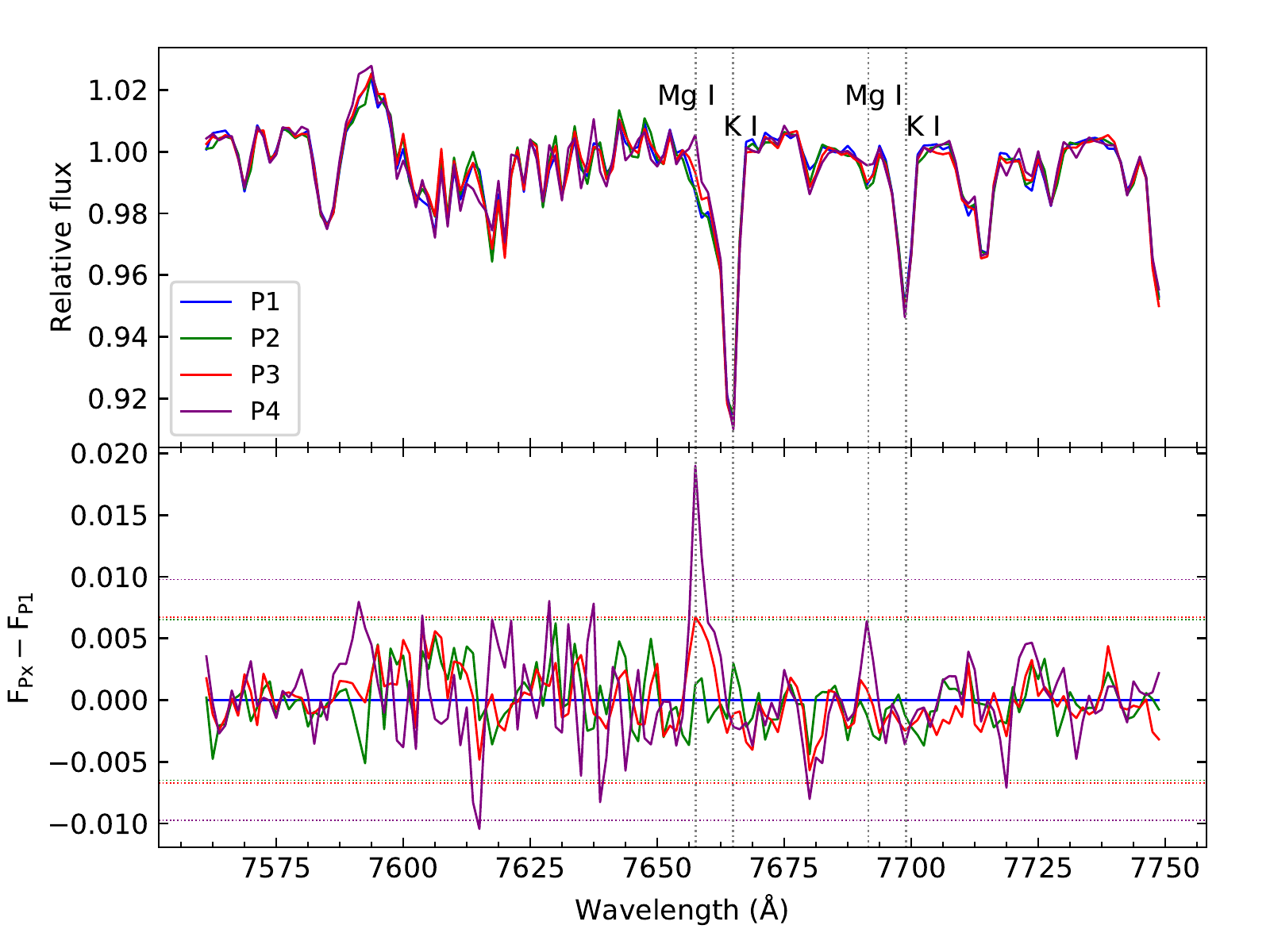}
 \includegraphics[width=\columnwidth]{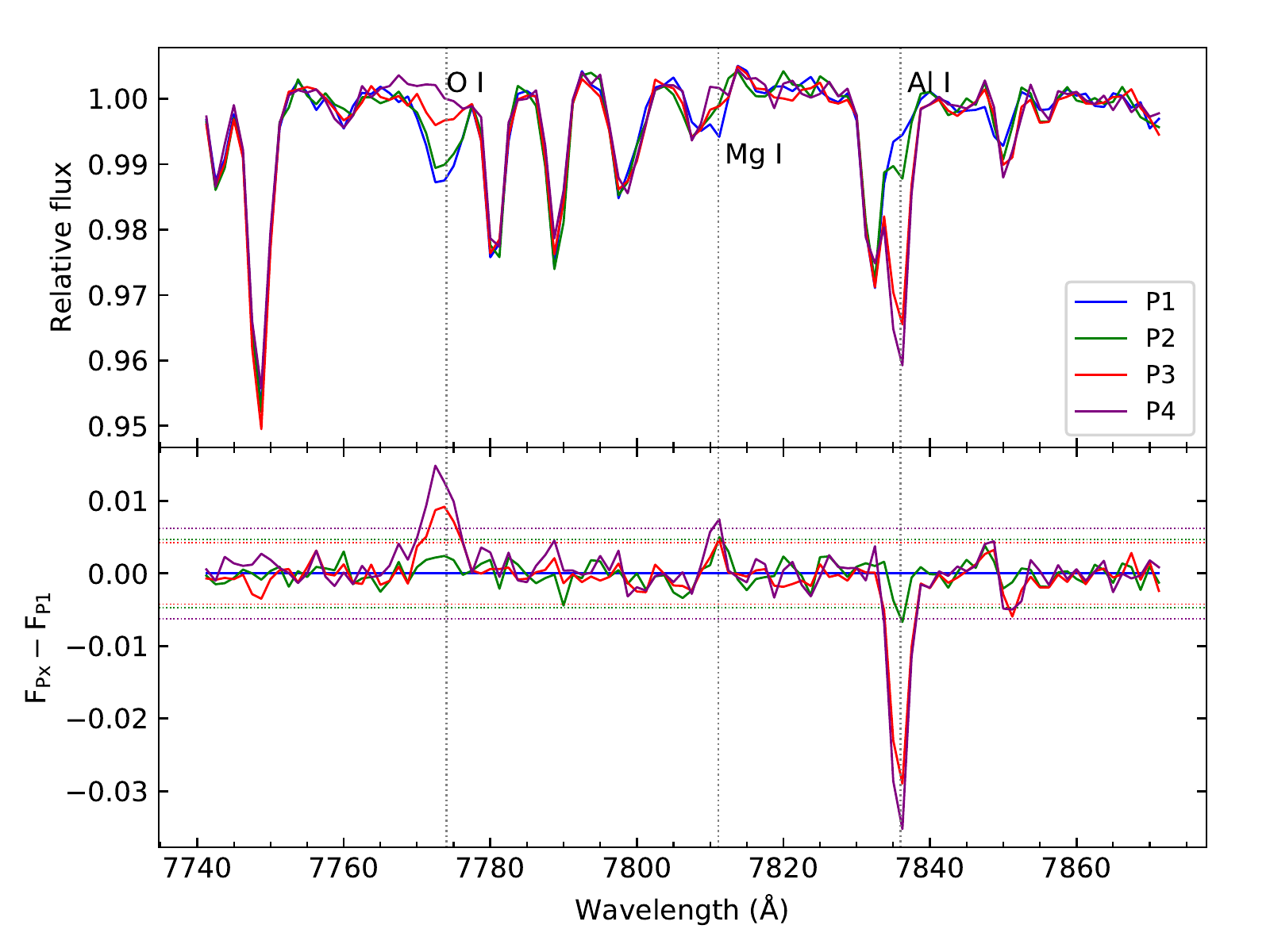}
 \includegraphics[width=\columnwidth]{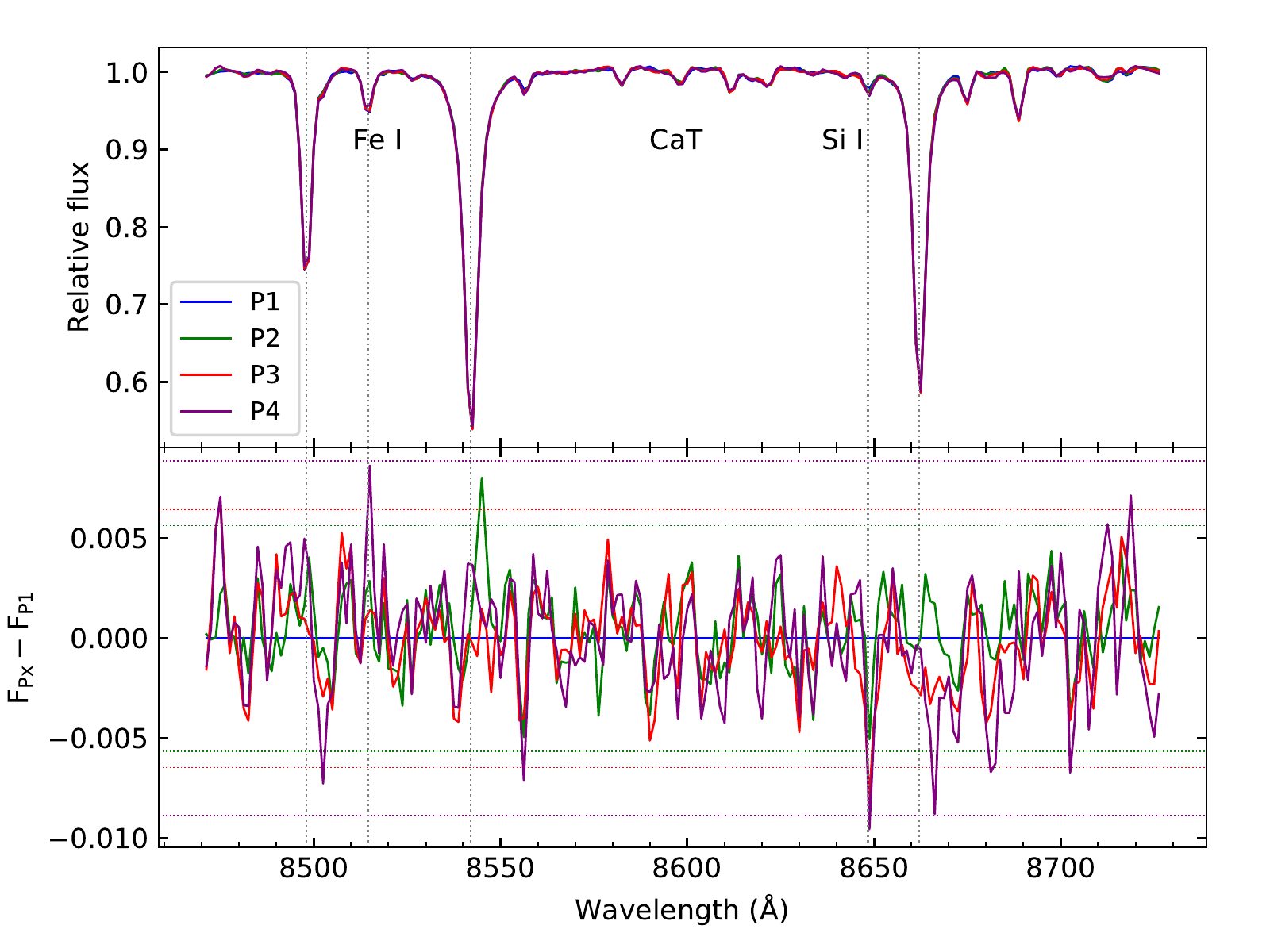}
 \includegraphics[width=\columnwidth]{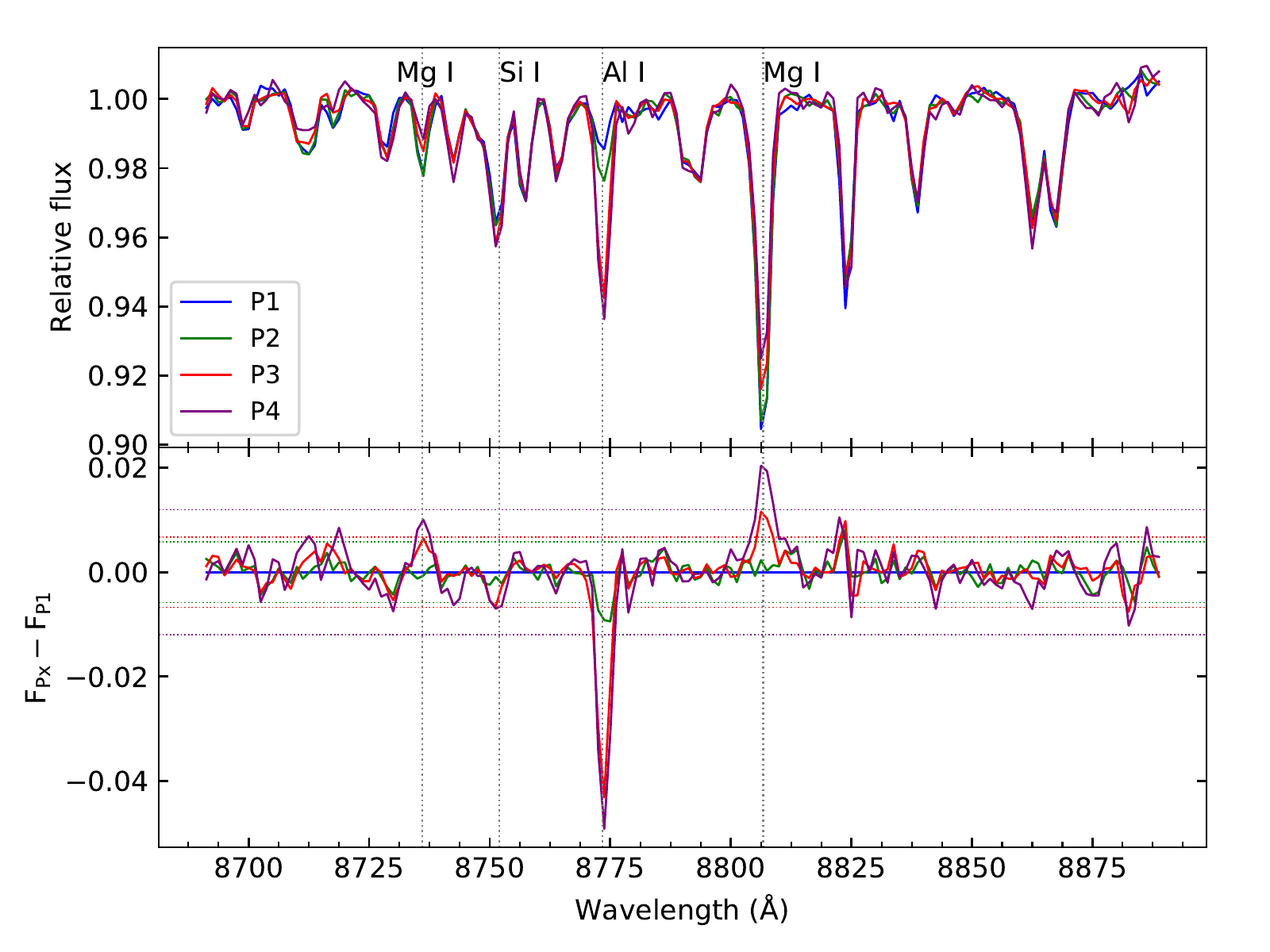}
 \caption{Same as Fig. \ref{spec_2808} for four additional spectral ranges.}\label{spec_2808_2}
\end{figure*}

\begin{figure*}[h!]
 \includegraphics[width=\columnwidth]{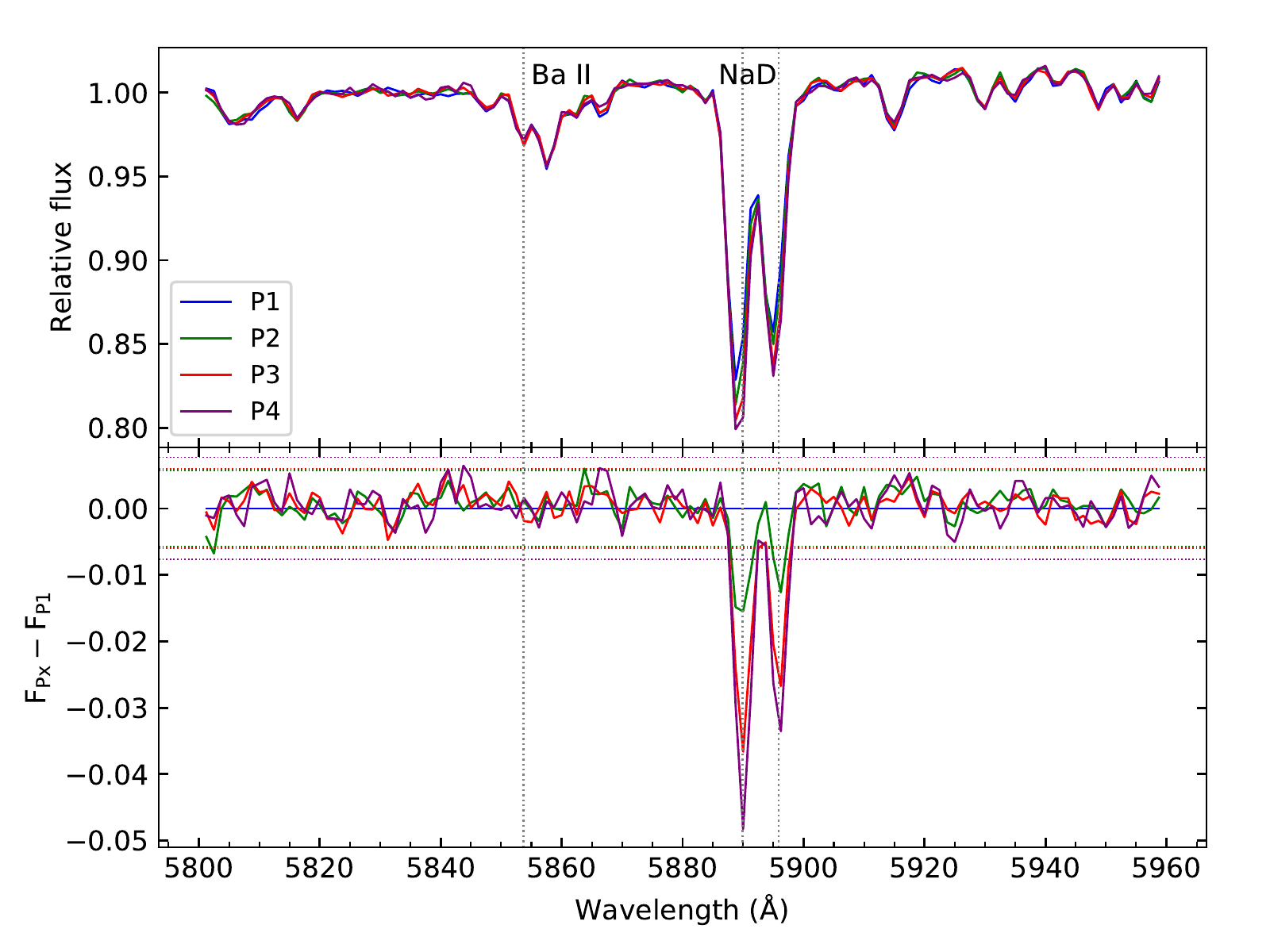}
 \includegraphics[width=\columnwidth]{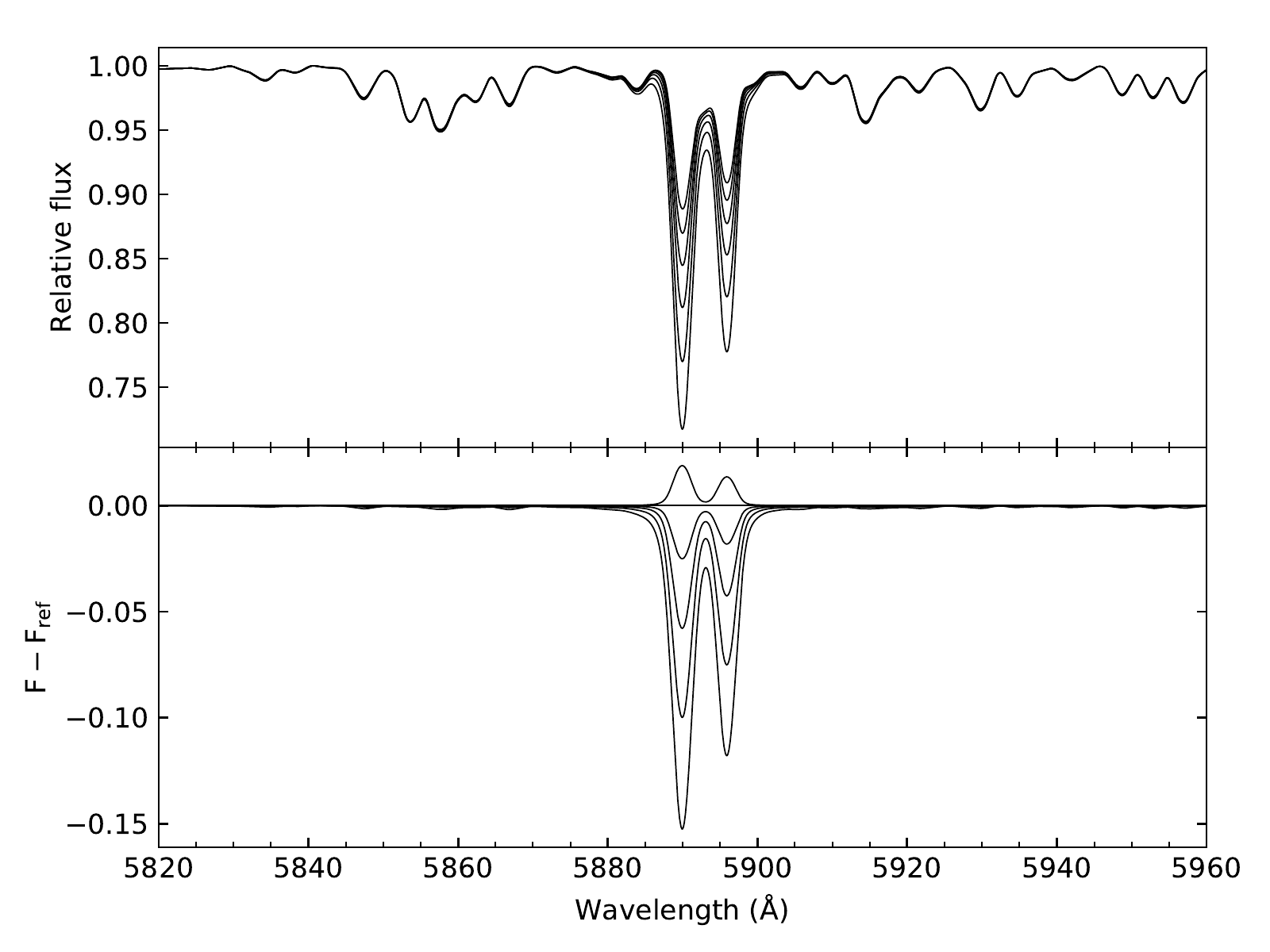}
  \caption{Left$-$ Same as Fig. \ref{spec_2808} and Fig. \ref{spec_2808_2} but for the NaD range. Right$-$ The NaD range in the synthetic spectra of P1 with [Na/M] varying from -0.25 to +1.00 by step of 0.25 dex. The residuals at the bottom are plotted with respect to the spectrum having [Na/M] = 0.00, as assumed for the primordial population.}\label{NaD}
\end{figure*}

We first searched for abundance variations by comparing the population spectra of \ngc, as well as those of other clusters, over the full MUSE wavelength range. This allowed us to identify spectral features that varied between the spectra. Table~\ref{llist} includes a list of lines for which we have detected variations, either in \ngc\ or in another cluster from our sample. The transitions marked with an asterisk were used in the quantitative analysis. 
The spectra were normalized by fitting the continuum over selected wavelength ranges, that are the same for each spectrum, and overplotted along with their flux differences (or residuals) in Figs. \ref{spec_2808} to \ref{NaD}. The differences in flux are always computed with respect to the spectrum of P1 ($\Delta F = F_{Px} - F_{P1}$). Along with the residuals we also indicated, for each population, a 3$\sigma$ confidence range, obtained from the (3$\sigma$-clipped) standard deviation of the flux differences over the wavelength interval. We included in Figs. \ref{spec_2808} and \ref{spec_2808_2} a selection of the most interesting spectral ranges showing some dramatic variations in Mg, Al and O. The quality of our combined spectra even allows us to see marginal differences in the residuals of a few weaker and/or strongly blended Mg lines. Over these wavelength ranges, we confidently detected differences of 1\% in relative flux. In the following subsections, we discuss our observations and abundance measurements for various elements.


\subsubsection{Na-O anticorrelation}
The most conspicuous sodium feature in the MUSE spectra are the NaD lines. Changes in the strength of these lines between the populations are clearly visible, with the sodium lines increasing in strength from P1 to P4 (Fig. \ref{NaD}). 
Even though the NaD doublet is blended with interstellar absorption, the residuals are peaking at the exact wavelength of the transition. This is also seen in our other GCs and indicates that we properly retrieve the photospheric variations.
We also observe a small variation in the \ion{Na}{i} $\lambda\lambda$5682, 5688 lines (see Fig.~\ref{spec_2808}) even though these lines are blended with other transitions of similar strength, notably from Fe and Si. 
As expected from the well established Na-O anticorrelation, the strength of the \ion{O}{i} ($\lambda$7774.2) line decreases from P1 to P4 (see Fig.~\ref{spec_2808_2}).

In order to translate the differences seen in these lines into abundances, we measured the EWs of the O and NaD residuals following the method described in Sect. 3.4 and used the synthetic population spectra described in Sect. 3.3.
We did not use the other Na doublet in our quantitative analysis because the residuals are weak. 
The synthetic spectra were first convolved with the MUSE line-spread function \citep{huss16} and the EWs of the lines of interest were computed in the same way as for the observed spectra. An example of the synthetic spectra computed for P1 with varying abundances of sodium is shown in the right panel of Fig. \ref{NaD}. To reproduce the EW differences measured from the observed spectra one must assume an abundance for the reference population (P1). For that we relied on the abundances measured by Carretta et al. (\citeyear{carr09,carr14}) as reported by \citet{mil15_2808}, in their Table\,2. Our P1 being equivalent to the population B of \citet{mil15_2808}, we set the sodium abundance of P1 to that of the cluster's metallicity (one tenth solar) and the oxygen abundance to [O/Fe] = +0.30. We considered a variation of $\pm$0.05 dex in the abundance of the reference population to obtain an uncertainty on its EW. 
For each line, the EW of that line at the P1 abundance was subtracted from the curve of growth of the remaining populations. These curves of growth, representing the EW differences as function of abundances, were then used to estimate the abundance differences of a given line, for each population. 

The resulting differential abundances are reported in Table\,\ref{Tabund} in the form of [Elem/Fe]$_{Px}$ - [Elem/Fe]$_{P1}$. As a reference, we also provide the abundance differences obtained from the values listed in Table\,2 of \citet{mil15_2808} when considering their population B as the reference population. Our abundances for sodium are in good agreement with the literature values. As for oxygen, the "bump" in the continuum of P4 required us to set the EW of the oxygen line in P4 to zero, meaning that the EW difference adopted for P4 was zero minus the EW of P1. Using the integral over the residual in that particular case would lead to an overestimated EW difference. 
Thus we could not constrain very well the oxygen variation in this population. Our lower limit for P4 is somewhat too high, due to the fact that the O line in our models do not disappear completely at low abundances ([O/M] = -1.0), although their EW is of comparable to the uncertainties ($\sim$13 m\AA). 

\subsubsection{Al-Mg anticorrelation}
\begin{table}
 \caption{Abundance differences between the populations of \ngc }\label{Tabund}
 \small
 \centering
 \begin{tabular}{l c c c }
 \toprule
 \toprule
 Element & Pop. & \multicolumn{2}{c}{$\Delta$ Abundance (Px - P1)} \\
  & 	& This work &\citet{mil15_2808} \\
 \midrule
 O 	& 2 & -0.11 $\pm$ 0.10 & -0.14 $\pm$ 0.09 \\
 O 	& 3 & -0.82 $\pm$ 0.35 & -0.67 $\pm$ 0.07 \\
 O 	& 4 & >$-$ 1.05 & -0.96 $\pm$ 0.14 \\
 \midrule
 Na	& 2 & 0.14 $\pm$ 0.06 & 0.18 $\pm$ 0.09 \\
 Na	& 3 & 0.35 $\pm$ 0.05 & 0.37 $\pm$ 0.10\\
 Na	& 4 & 0.50 $\pm$ 0.06 & 0.76 $\pm$ 0.14 (0.60) \\
 \midrule
 Mg	& 2 & -0.03 $\pm$ 0.02 & 0.03 $\pm$ 0.14 \\
 Mg	& 3 & -0.15 $\pm$ 0.06 & ... (-0.18) \\
 Mg	& 4 & -0.20 $\pm$ 0.09 & ... (-0.43) \\
 \midrule
 Al	& 2 & 0.18 $\pm$ 0.17 & 0.18 $\pm$ 0.15 \\
 Al	& 3 & 0.87 $\pm$ 0.16 & ... (1.00) \\
 Al	& 4 & 1.12 $\pm$ 0.16 & ... (1.20) \\
 
  \bottomrule                               
\end{tabular}
\tablefoot{ The values in parenthesis are estimated from the abundances of \citet{carr15_2808}. }
\end{table}

The magnesium $b$ triplet and \ion{Mg}{i} $\lambda$8806.75 are the two diagnostic features for Mg variations in the MUSE spectra. Magnesium is clearly depleted in P3 and P4 and the residuals are well above the 3$\sigma$ limit. On the other hand we do not detect Mg variations between P1 and P2. 
The variations in aluminum are anti-correlated with those of magnesium. The Al lines increase in strength from P1 to P4 with a steeper increase between P2 and P3. This ``bimodality'' in aluminum abundances was also observed by \citet{carr09} (see their Fig. 6) in the dozen RGB stars for which they derived abundances of these elements. 

\begin{figure}
 \includegraphics[width=\columnwidth]{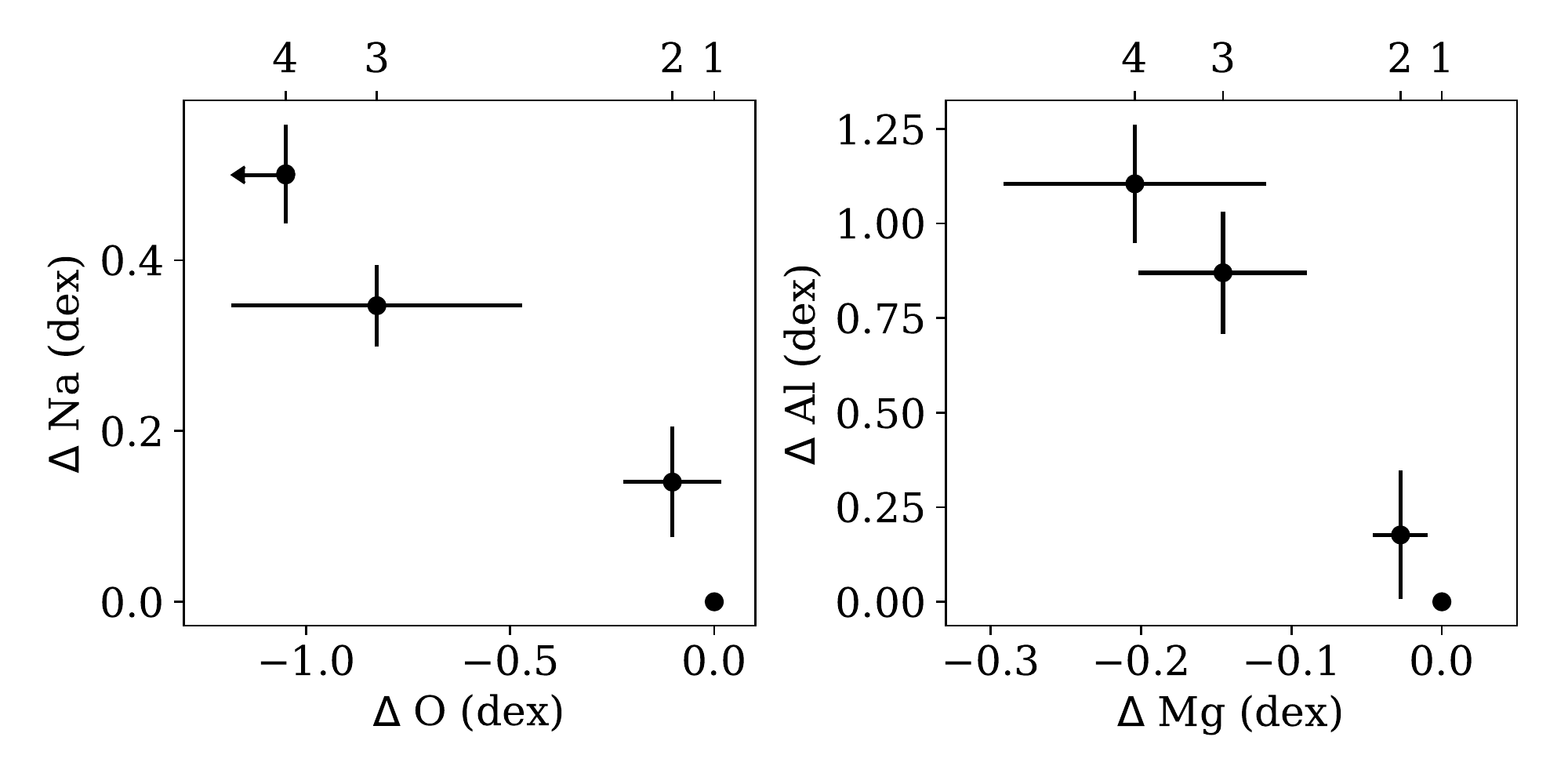}
 \caption{Anticorrelations between the Na-O (left) and Al-Mg (right) abundances. The population numbers are indicated on the top axes. }
 \label{anti-corr} 
\end{figure}

The quantitative analysis was performed similarly to that of the Na and O lines. 
For magnesium, we used three regions to compute EWs, a first region including the two bluest lines of the Mg $b$ triplet ($\lambda\lambda$5167, 5173) and the other two regions including the third Mg $b$ line ($\lambda$5183) and the line at 8806.8 \AA.  The Mg abundance of P1 was set to [Mg/M] = +0.38 and the Al abundance of P1 to that of the cluster's metallicity.
For aluminum, we used the four spectral lines displayed in Fig. \ref{spec_2808} and \ref{spec_2808_2}. The differential abundances obtained from the various lines are plotted in Fig. \ref{abund_diff} and the values reported in Table \ref{Tabund} were obtained from the average of the different lines. The abundance differences reported in Table \ref{Tabund} are also illustrated in Fig. \ref{anti-corr}.
In \citet{mil15_2808}, only stars from their populations B and C (corresponding to our P1 and P2) have Al and Mg abundances. For P3 and P4, we used instead the average abundances listed in Table 7 of \citet{carr15_2808}, using the I1 and I2 populations as equivalent of our P3, and the E population for our P4.  
Although the individual abundance values obtained from the different lines are not perfectly consistent (Fig.~\ref{abund_diff}), the trend across the populations is similar and the average abundances are in good agreement with the literature values. Only our abundance variation of Mg in P4 does not reach the depletion of $\sim-$0.4 dex found by \citet{carr15_2808}.

\citet{carr15_2808} also observed an anticorrelation between the Si and Mg abundances, but with a relatively small variation in the silicon abundances ($\sim$0.2 dex). This is much smaller than that of the other elements (Al, Mg, O, and Na) and thus more difficult to observe in our spectra. Few silicon lines are strong or relatively well isolated in our spectra, but small features in the residuals could be identified with two Si lines ($\lambda\lambda$8646, 8752). Their behavior is as expected with the lines being stronger in P3 and P4 (see Fig.~\ref{spec_2808_2}).

\subsubsection{Elements without variations}

Three of the investigated atomic elements do not show variations between the populations: K, Ca, and Ba. 
Star-to-star scatter in potassium abundances are normally not observed in GC \citep{tak09,mucc17} but, interestingly, \ngc\ is one of the two clusters (along with NGC\,2419) where the K abundance has been observed to anti-correlate with O and Mg while correlating with Na and Al \citep{mucc15}. However the K abundance variation is relatively small ($\sim$ 0.25 dex) and the correlation with Mg not particularly strong (see Fig. 12 of \citealt{carr15_2808}). In the MUSE spectra, the K resonance lines are blended with their interstellar components as indicated by the blue-shifted position of the observed K lines in Fig.~\ref{spec_2808_2} (such a shift is also visible in the NaD lines). The potassium lines are also close to strong telluric absorption that is not perfectly accounted for in our telluric models. This discrepancy explains the feature seen on the blue side of \ion{K}{i} $\lambda$7664 (see also Fig. 6 of \citealt{huss16}). These are not ideal conditions to detect small variations as those expected for the potassium lines. It is nevertheless worth pointing out that despite the residual telluric features, we detect variations in the Mg line at 7659~\AA\ (see Fig.~\ref{spec_2808_2}). 

We do not see variations in the residuals of the Ca triplet (CaT) lines, which is in agreement with the results of \citet{carr15_2808} who reported no variations in Ca abundances related with the populations. 
Finally, while barium shows abundance variations in a handful of GC, this does not appear to be the case in NGC\,2808 as none of the Ba lines we inspected display variations. We did not find literature on the Ba abundances of RGB stars in this cluster, however, barium abundances were measured in HB stars and found to be consistent with a constant value \citep{mar14}. 

One last interesting feature is the H$\alpha$ line, for which the residual detected in P4 is consistent with the previous observations (see Sect. 4.1) that this population contains hotter stars, since the Balmer lines become stronger with increasing \teff. Because they are believed to be enhanced in helium, the stars belonging to P4 are expected to be not only hotter, but also to have slightly lower surface gravities (by 0.05 dex). We made sure that this would not significantly change our abundance determinations by re-computing the synthetic spectra of P4 using a lower surface gravity. The abundances derived with these "lower log $g$" spectra were differing only by 0.01$-$0.02 dex compared to the results listed in Table~\ref{Tabund}. 
For the three other populations, their H$\alpha$ lines are remarkably similar (Fig~\ref{spec_2808}). In fact, the whole spectra of the four populations are extremely similar, besides the lines affected by abundance variations, indicating that they are minimally affected by the differences in their underlying distributions in \teff\ and log $g$.

\begin{figure}
\centering
 \includegraphics[width=0.9\columnwidth]{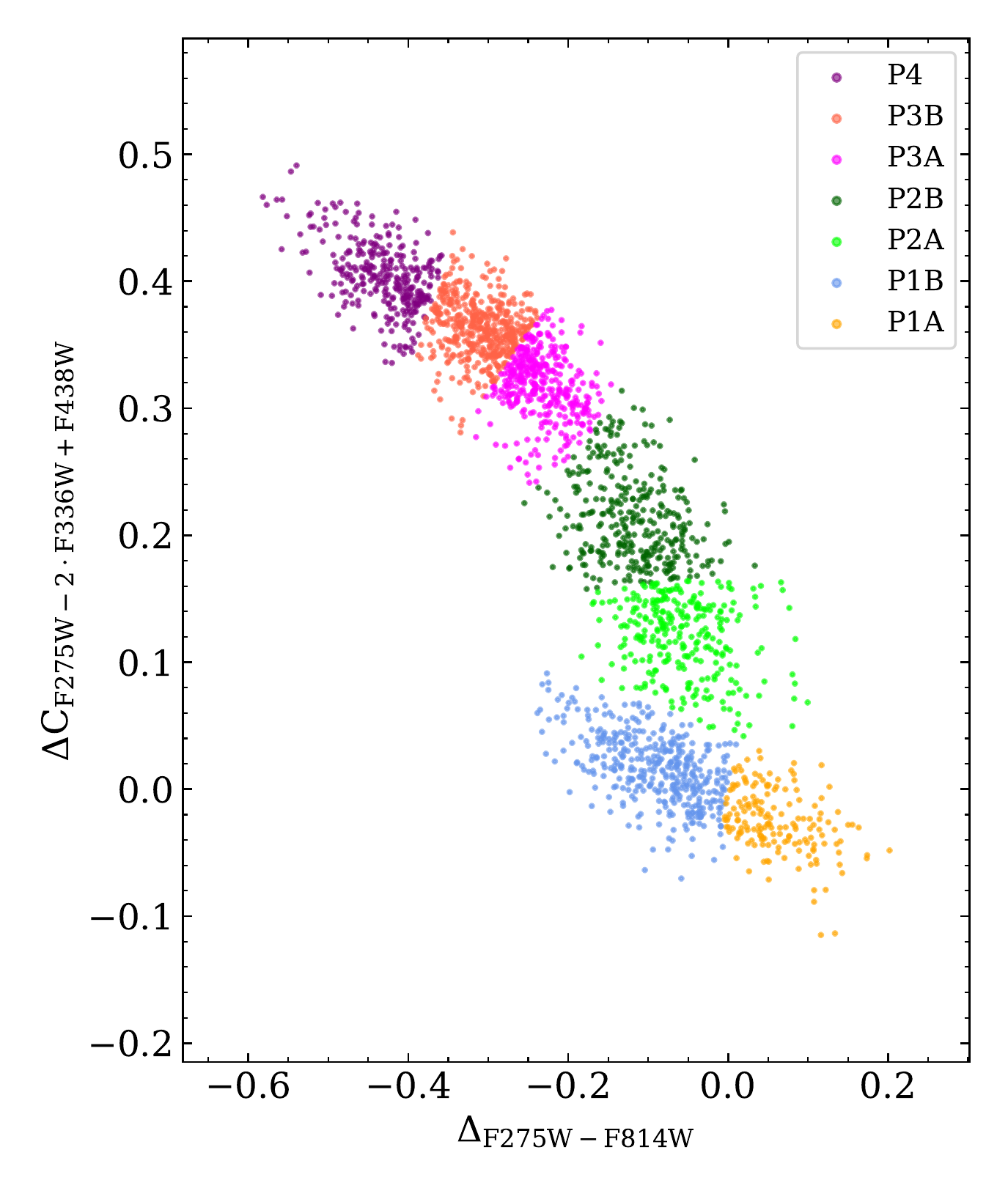}
 \caption{Chromosome map of \ngc\ including additional subdivisions. The population 1 is divided in P1A and P1B following the nomenclature of \citet{mil15_2808}. Populations 2 and 3 are also each separated in two subgroups. The population groups are listed in the legend from the group on top of the \cmap\ (P4) to that at the bottom (P1A). 
 }
 \label{cmap_AB} 
\end{figure}

\subsection{A closer look at the primordial population}

\begin{figure*}[t]
\sidecaption
\includegraphics[width=12cm]{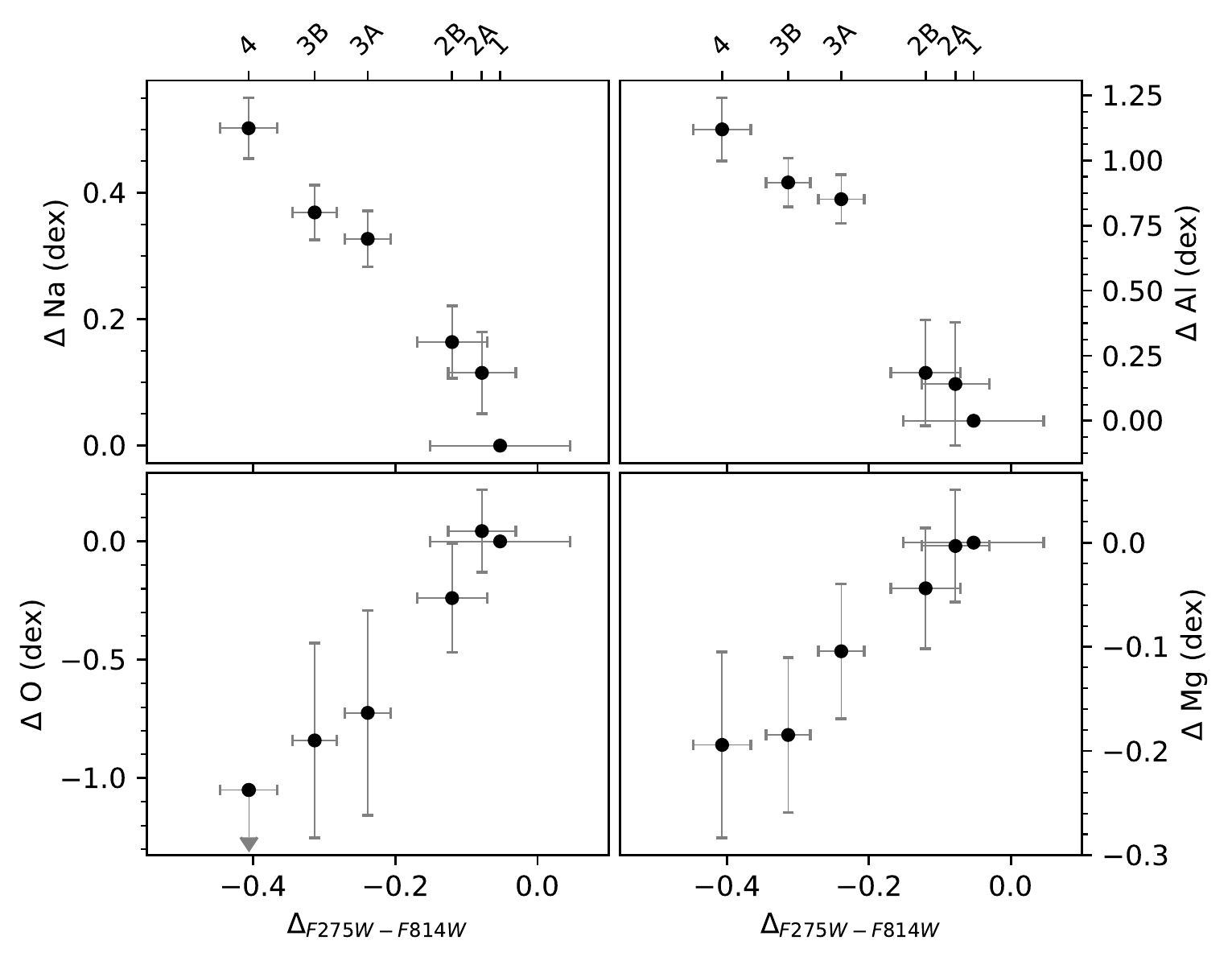}
\caption{Abundance variations measured for O, Na, Mg and Al plotted against the median pseudo-color $\Delta_{F275W-F814W}$ of the populations. The errors on the x-axis represent the 1$\sigma$ dispersion in peusdo-color distribution of the given population. The population numbers are indicated on the top axes. }
\label{abund_pop}
\end{figure*}

\citet{mil15_2808} identified five populations based on the \cmap\ and CMDs of NGC\,2808. The additional population comes from a subdivision of population\,1 stars in two groups that they identified as Population A and B. Figure \ref{cmap_AB} shows our \cmap\ updated with this additional subdivision among the P1 stars. Indeed some GCs, such as \ngc, have a P1 that is rather extended along the x-axis ($\Delta_{F275W - F814W}$), while for some other clusters the extension is much more restricted \citep{mil17}. Comparisons between observed and synthetic photometry suggested that the extent of the P1 stars is mostly due to variations in helium abundances (with He increasing toward the left in \cmaps; \citealt{mil18,lar18}). However such He variations in stars that would otherwise have a rather homogeneous composition is challenging to explain via the current enrichment mechanisms proposed in the literature \citep{bas18}. An alternative possibility suggested to explain the photometric spread of P1 stars is variations in iron abundance \citep{mil15_2808,marino19}. 
In the case of \ngc, it is unfortunate that no star from the \citet{carr15_2808} sample could be associated with the Population A stars of \citep{mil15_2808}, thus spectroscopic information was lacking concerning that intriguing group of stars. It is only recently that \citet{cab19} measured abundances of light elements in six stars found along the P1 of \ngc\ (including three likely belonging to PA) and found the abundances to be homogeneous among the six objects.

Given the state of the current debate concerning the cause of the P1 spread in the \cmap\ and the rather mysterious status of the population A stars in \ngc, we explored this further with the MUSE spectra. We created spectra for the population 1A (including 79 stars) and 1B (189 stars) and compared them as we did for P1 to P4. The resulting spectra and residuals are presented in Fig. \ref{spec_AB}. As expected from the recent findings of \citet{cab19}, we do not detect significant variations in any of the spectral lines investigated. 
Measurement of the EW differences between P1A and P1B and their comparison with synthetic model spectra, as done in the previous subsection, confirmed that the differences are consistent with no abundance variations. By considering the uncertainties on the measured EWs ($\sim$16 m\AA), we estimated the abundance variations between P1A and P1B to be: $-$0.08 $\pm$ 0.2 dex for O, 0.03 $\pm$ 0.07 dex for Na, 0.04 $\pm$ 0.23 dex for Al and 0.02 $\pm$ 0.05 dex for Mg. 
This supports the hypothesis that the color variation among the P1 stars could be related to helium, whose abundance is notoriously difficult to quantify via spectroscopic analysis. As for the possible presence of an iron spread among the population 1 stars, it will be discussed in Husser et al. (2019; submitted to A\&A).

\subsection{Abundance variations across the \cmap.}

To investigate the possibility of abundance variations across the \cmap\ and within populations, we further divided P2 and P3 in two subgroups as indicated in Fig.~\ref{cmap_AB}. For these four additional populations, we created population spectra, measured EW differences and translated these EW differences into abundance differences following the method described in the previous sections. 
We used again P1 as reference population. 
We show the resulting abundances in Fig. \ref{abund_pop}. The abundances are plotted with respect to the median pseudo-color $\Delta_{F275W-F814W}$ (x-axis on the \cmap) of each population, with the errorbars on $\Delta_{F275W-F814W}$ representing the standard deviation of the distribution in pseudo-color. 

Although the uncertainties on O and Mg abundances are large, a gradual trend in these elements abundances is seen when moving across the \cmap. We note however that for these two elements, the abundances of the P2A group appear to be the same as that of the primordial (P1) population. 
This behavior in \ngc\ is similar to what is seen in Fig.~13 of \citet{marino19}; the stars at the "bottom" of P2 (sharing a similar $\Delta_{F275W-F814W}$ than the P1 stars) have similar abundances in Mg and O. 

Na and Al abundances display an increasing trend when moving to the left of the \cmap. 
We see a drastic increase in the abundance of Al between the 2B and 3A populations. This further supports the presence of a gap in Al abundances between P2 and P3. We recall that this is also seen in the abundances measured by Carretta et al. (\citeyear{carr15_2808,carr18}) where there is a difference of $\sim$0.7 dex between the Al abundance of their P2 and I1 groups. As for the trend in sodium, there is also some evidence of discreet changes between the populations.

\section{Conclusion }

We used the spectra of more than 1100 RGB stars in \ngc\ collected with the MUSE integral field spectrograph to look at the abundance variations among the multiple stellar populations of the cluster. We used the pseudo two-color-diagram (or \cmap) to optimize the separation of the stars in their respective populations according to the classification presented in \citet{mil15_2808}. Because of their low resolution, the MUSE spectra are not optimal for performing abundance analysis on individual stars. Therefore, we followed a different approach and combined the spectra of all stars within a given population to obtain one high S/N spectrum per stellar population. By comparing the spectra of the different populations, we readily detected variations in the spectral lines of Na, O, Al, and Mg following the expected Na-O and Al-Mg anticorrelations. We measured equivalent widths for a set of spectral lines to perform a differential abundance analysis and estimated abundance differences between the populations.
Our overall results for the abundance variations of Na, O, Al, and Mg compare well with what is expected from literature. Interestingly, we found a sharp variation in aluminum abundance between two of the populations (P2 and P3). 
Considering that we worked with low-resolution spectra, used a different approach, and different spectral lines, than Carretta et al. (\citeyear{carr06,carr15_2808,carr18}) to make our measurements, we find the agreement quite remarkable.

We also examined the properties of the stars belonging to two subgroups (P1A and P1B) forming the primordial population (P1). Based on the photometric properties of these stars, previous investigations suggested that the extension of the P1 stars in \cmaps\ can be explained by variations in helium abundances \citep{mil15_2808,lar18}. Our investigation of the P1A and P1B stars did not reveal significant variations in O, Na, Al, or Mg, indicating that these elements have homogeneous abundances. Our findings are in line with the recent work of \citet{cab19} who did not detect variations in the abundances of light elements among the primordial population, although their sample included only six objects.

Finally, even though our method does not provide abundances as precise as those obtained from high-resolution spectroscopy of individual stars, we can obtain reliable estimates of the abundance variations between populations. 
But most importantly, the sheer amount of spectra collected in the last years as part of the MUSE GC survey allows us to readily detect variations in line strength between different populations spectra. This also provides flexibility in terms of defining populations and searching for abundance variations accross the chromosome maps. 
Following the method used for \ngc\ in this paper, we will further explore the chemical properties of the RGB stars in other clusters included in the MUSE survey that present an interesting set of populations, such as NGC\,7078, NGC\,1851, and $\omega$ Centauri.

\begin{acknowledgements}
 We would like to thank I. Hubeny for his sharing his latest version of SYNSPEC. We also thank D. Nardiello for sharing with us the ID matches between the HUGS and ACS catalogs.
 We acknowledge funding from the Deutsche Forschungsgemeinschaft (grant DR 281/35-1 and KA 4537/2-1) and from the German Ministry for Education and Science (BMBF Verbundforschung) through grants 05A14MGA, 05A17MGA, 05A14BAC, and 05A17BAA.
SK and NB gratefully acknowledge financial support from the European Research Council (ERC-CoG-646928, Multi-Pop). NB also gratefully acknowledges financial support from the Royal Society (University Research Fellowship).
JB acknowledges support by FCT/MCTES through national funds by grant UID/FIS/04434/2019 and through Investigador FCT Contract No. IF/01654/2014/CP1215/CT0003.
This work has made use of the VALD database, operated at Uppsala University, the Institute of Astronomy RAS in Moscow, and the University of Vienna. This research has made use of NASA's Astrophysics Data System. 

\end{acknowledgements}



\bibliographystyle{aa}


\begin{appendix} 

\section{Additional Figures}

\begin{figure*}
\centering
 \includegraphics[width=\columnwidth]{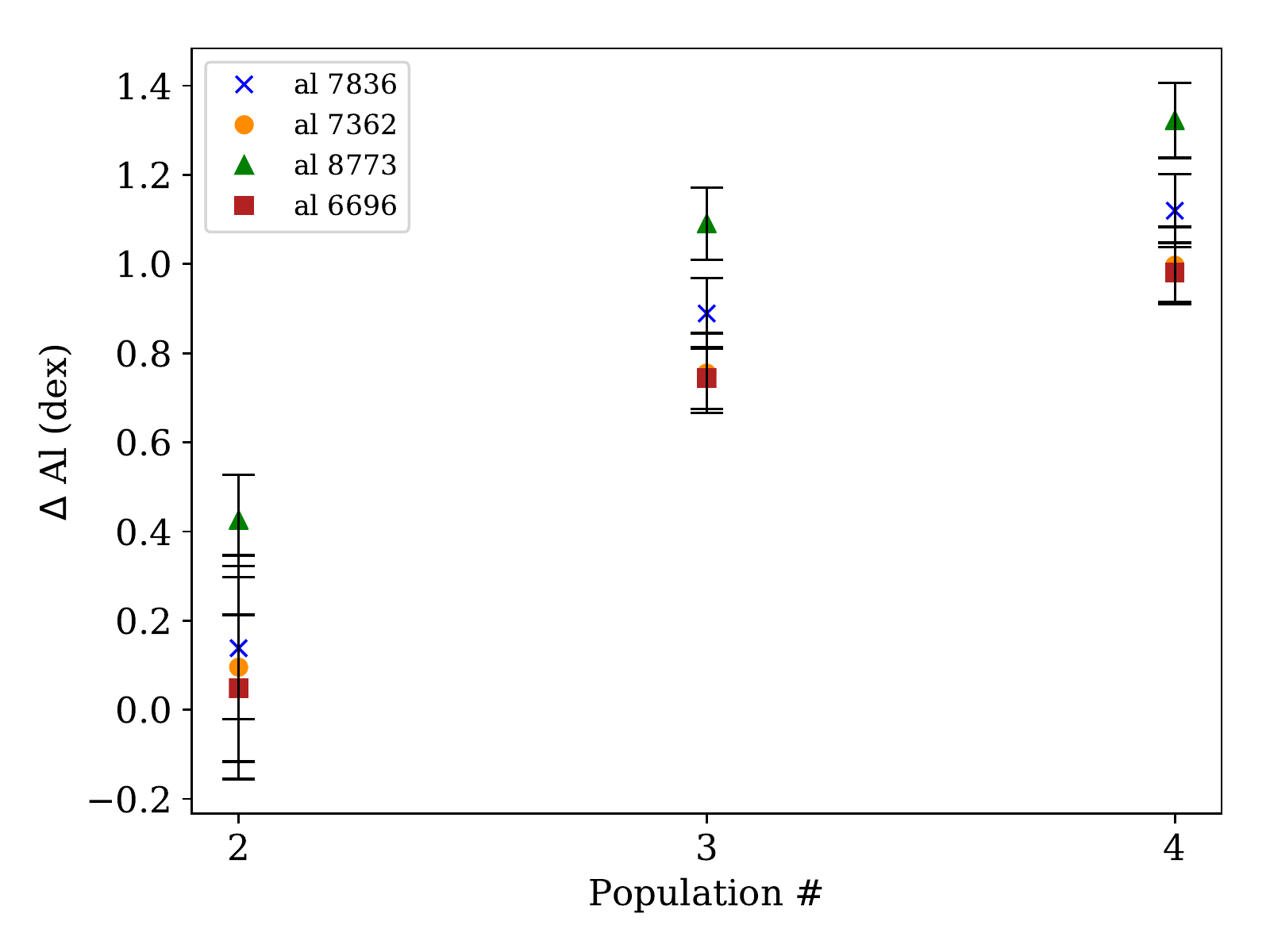}
  \includegraphics[width=\columnwidth]{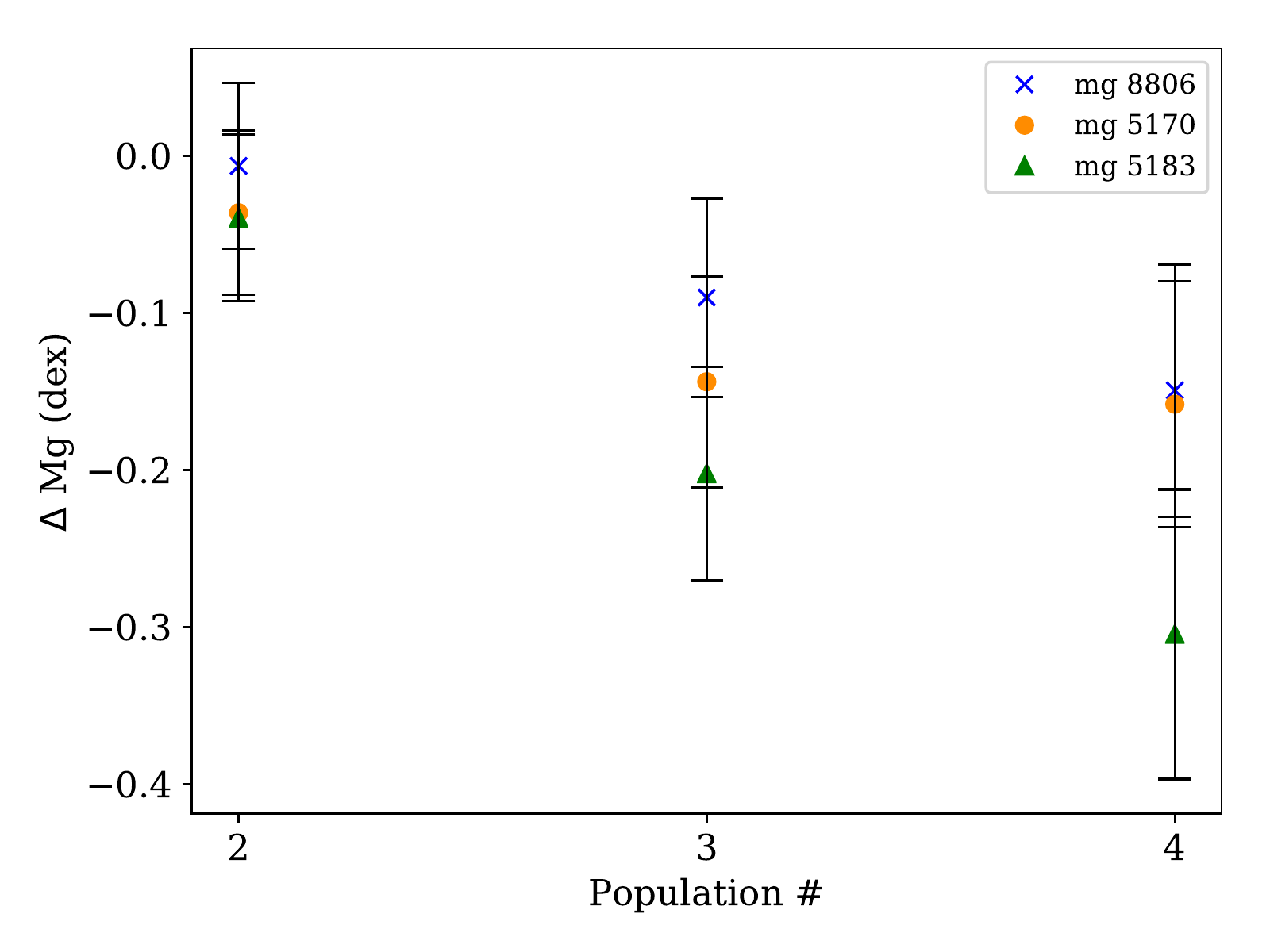}
 \caption{Abundances differences obtained from the Al (left panel) and Mg (right panel) lines for the three populations. The abundances are expressed as [X/Fe]$_{Px}$ - [X/Fe]$_{P1}$. }
 \label{abund_diff}
\end{figure*}

\begin{figure*}[t]
 \includegraphics[width=\columnwidth]{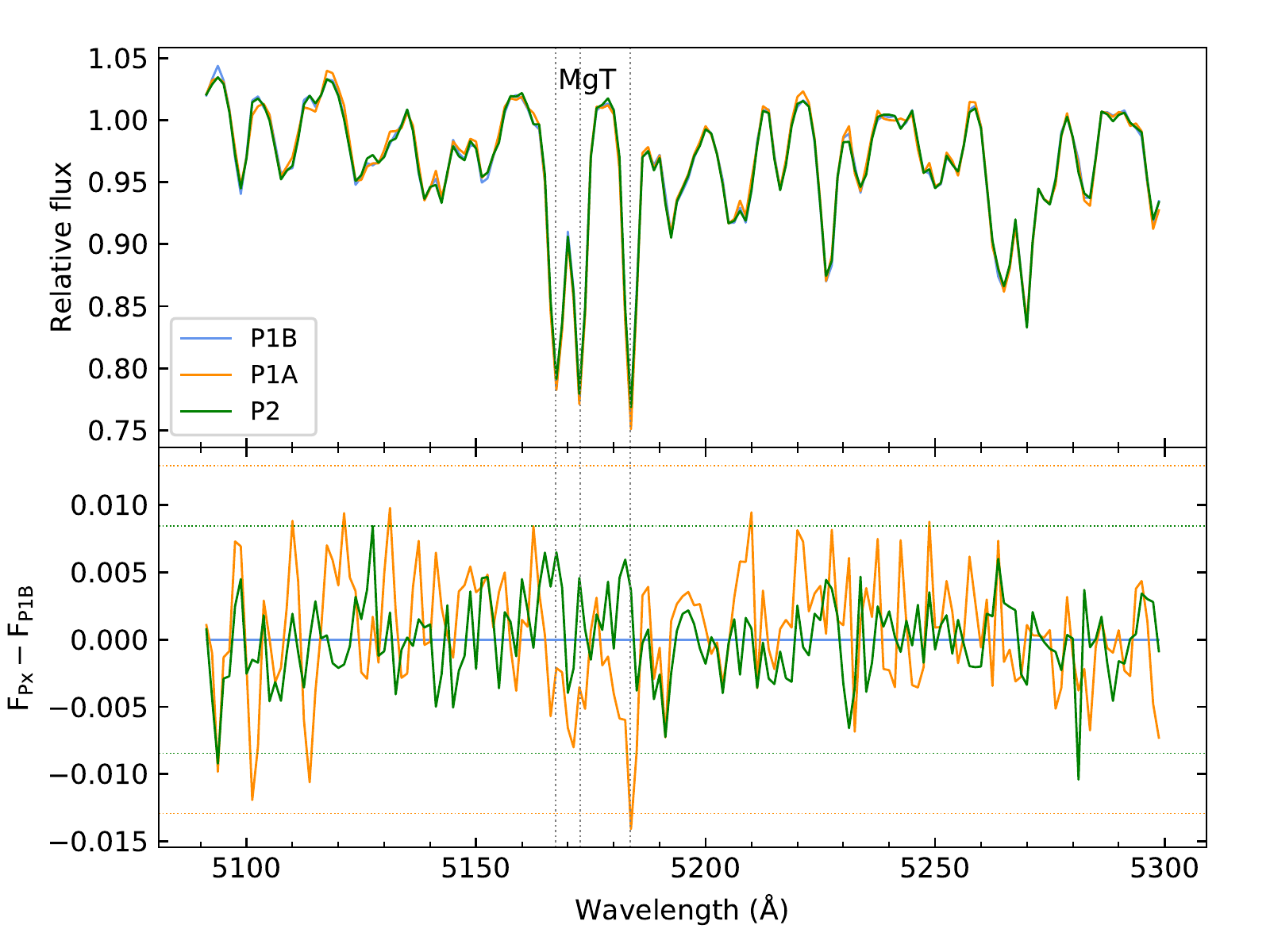}
 \includegraphics[width=\columnwidth]{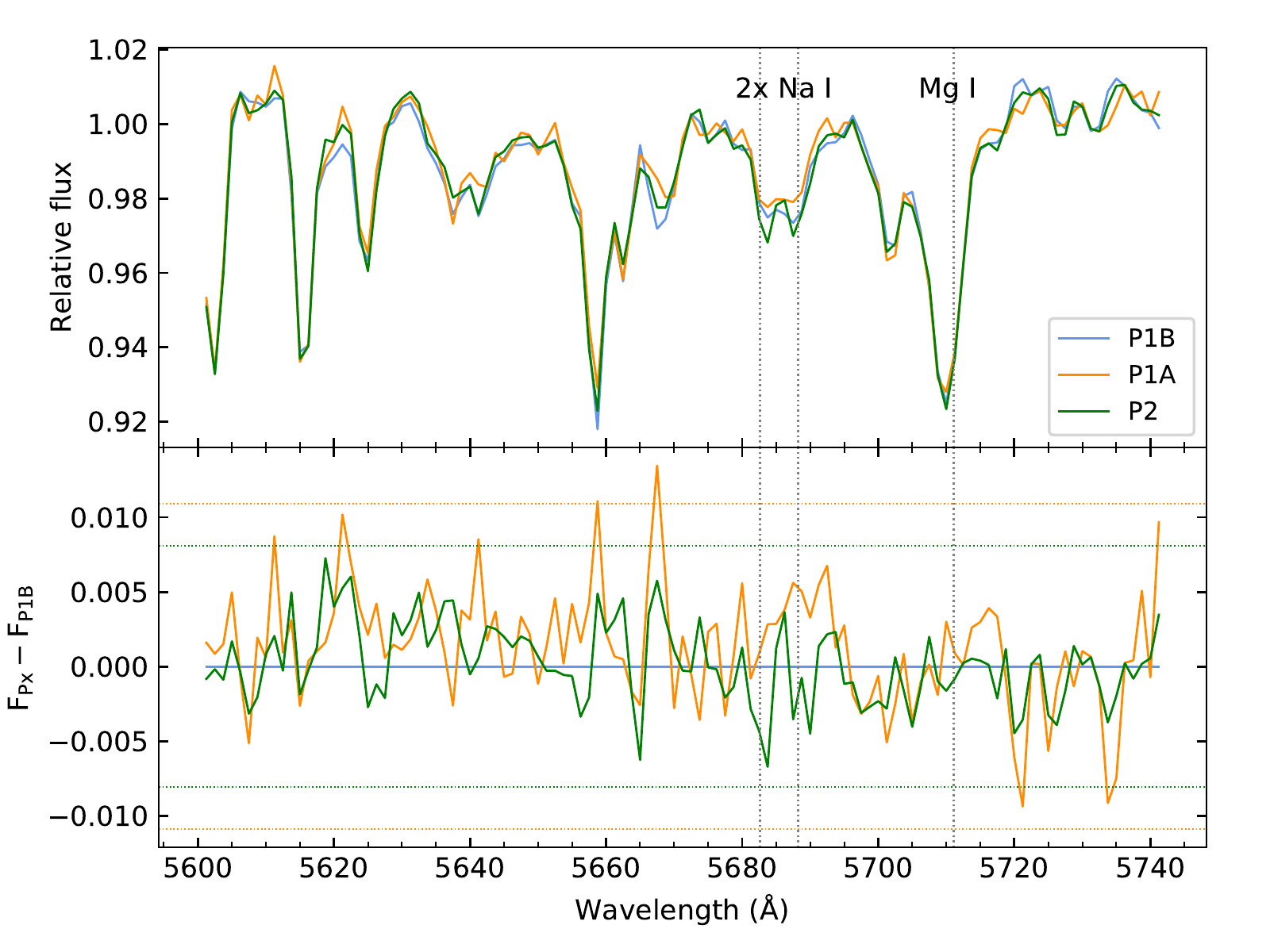}
 \includegraphics[width=\columnwidth]{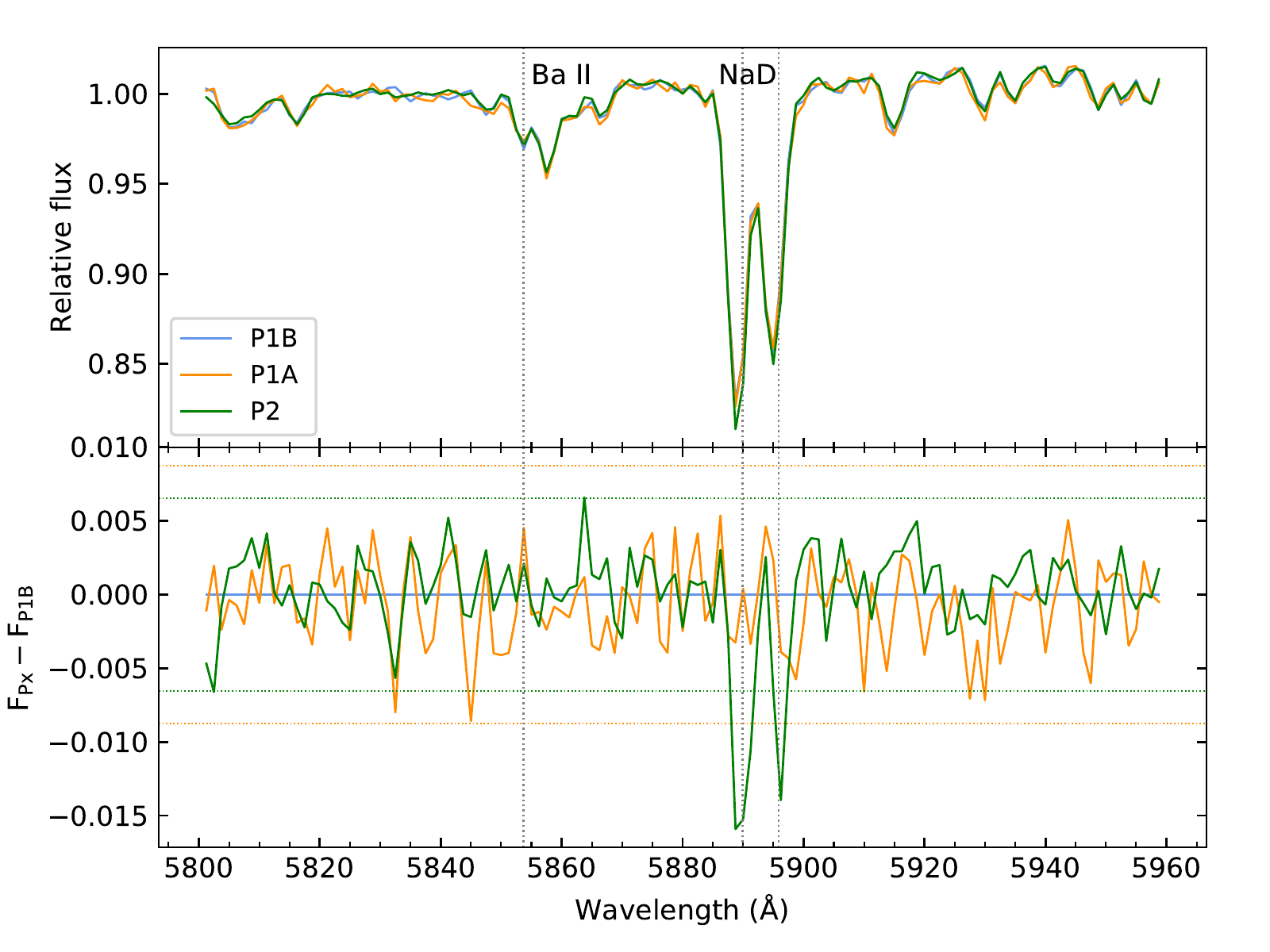}
 \includegraphics[width=\columnwidth]{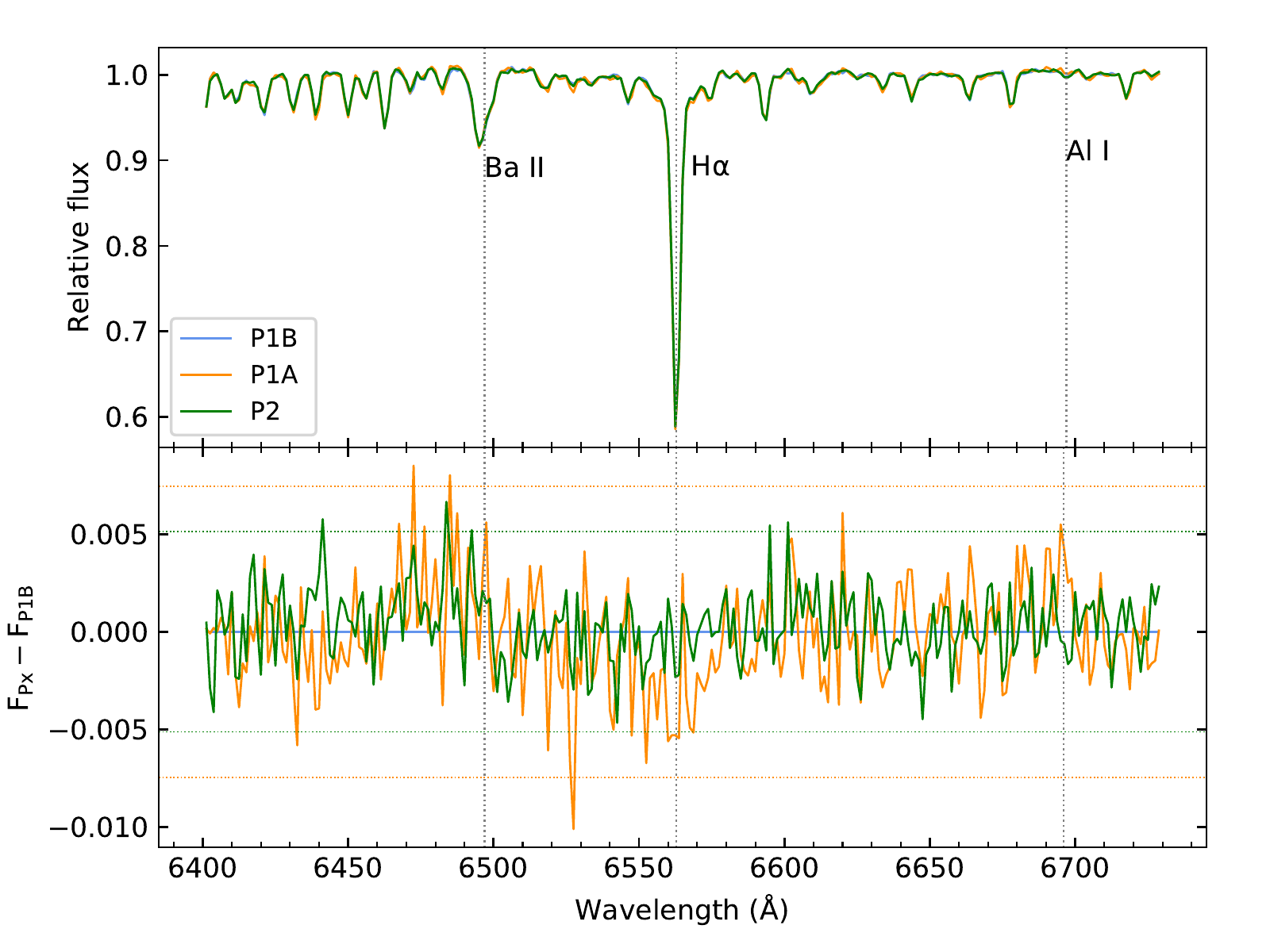}
 \includegraphics[width=\columnwidth]{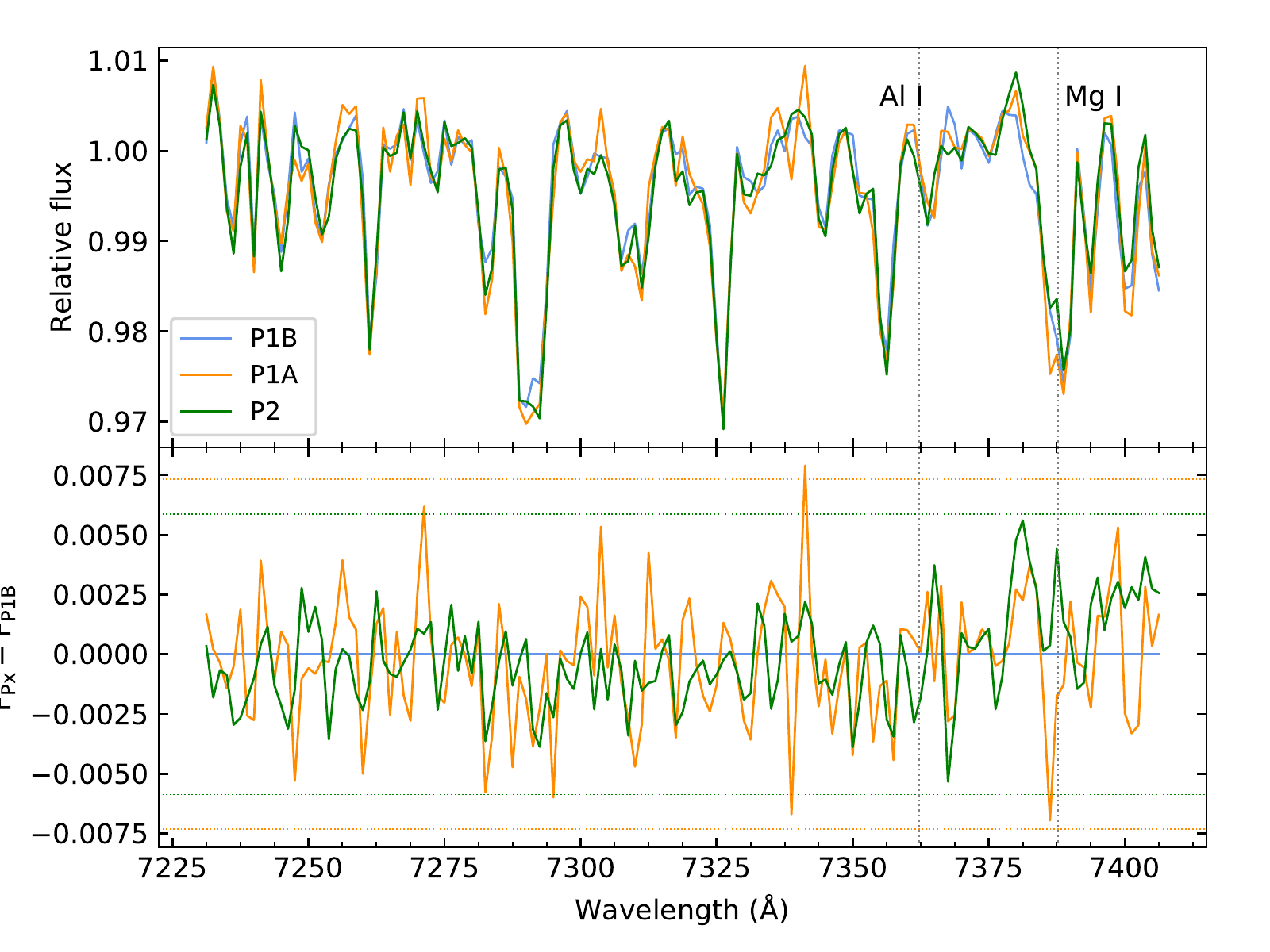} .
   \caption{Comparisons between the spectra of the population A and B in NGC\,2808. As a reference we also show the spectrum of P2. 
   The residuals on the bottom panels are plotted as the difference between the flux of population 1A (or 2) and 1B (F$_{Px}$ - F$_{P1B}$). The horizontal dotted lines represent the 3$\sigma$ value of the residuals. }
  \label{spec_AB}
  \end{figure*}
 \addtocounter{figure}{-1}
 \begin{figure*} 
  \includegraphics[width=\columnwidth]{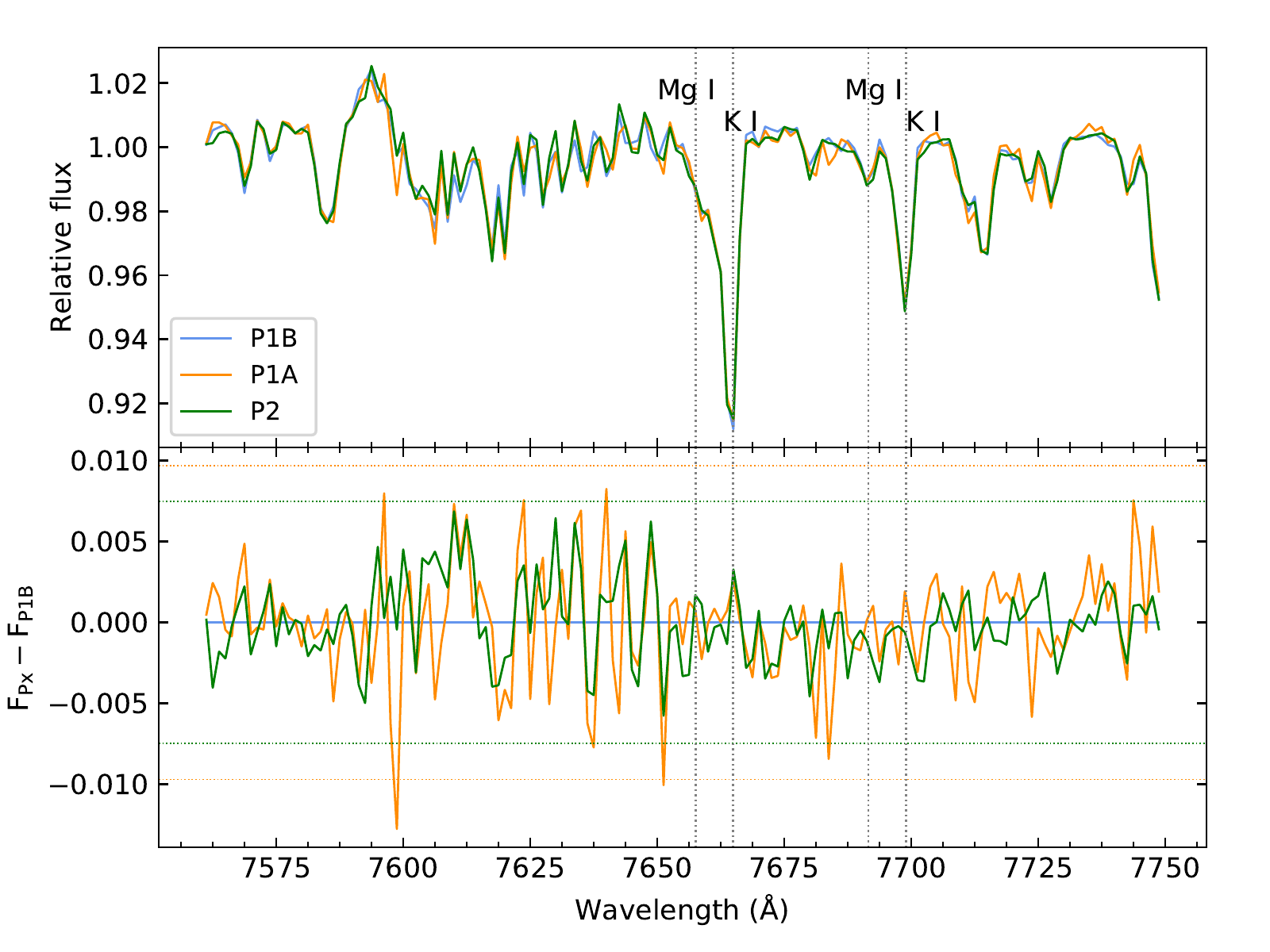}
 \includegraphics[width=\columnwidth]{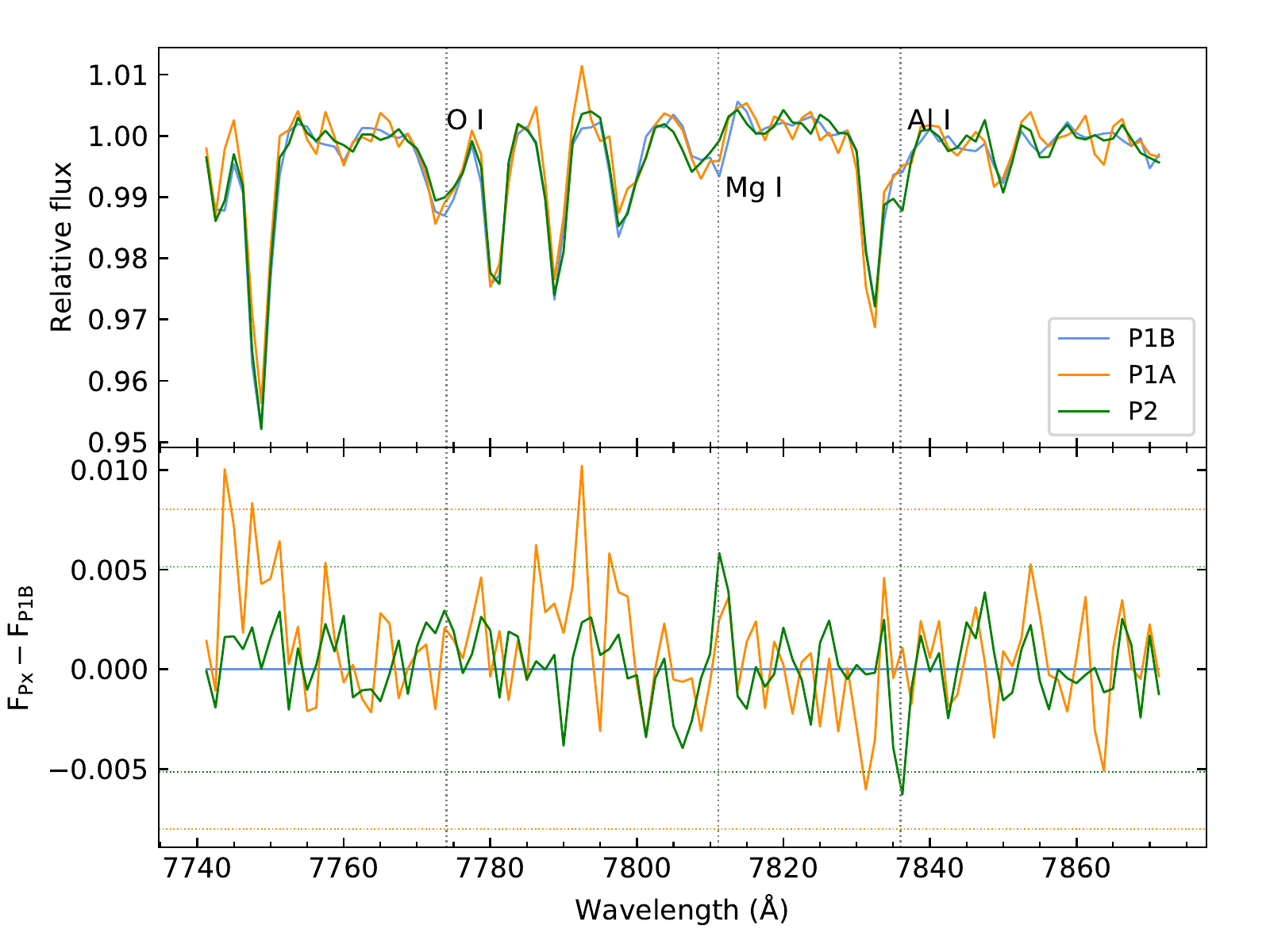}
 \includegraphics[width=\columnwidth]{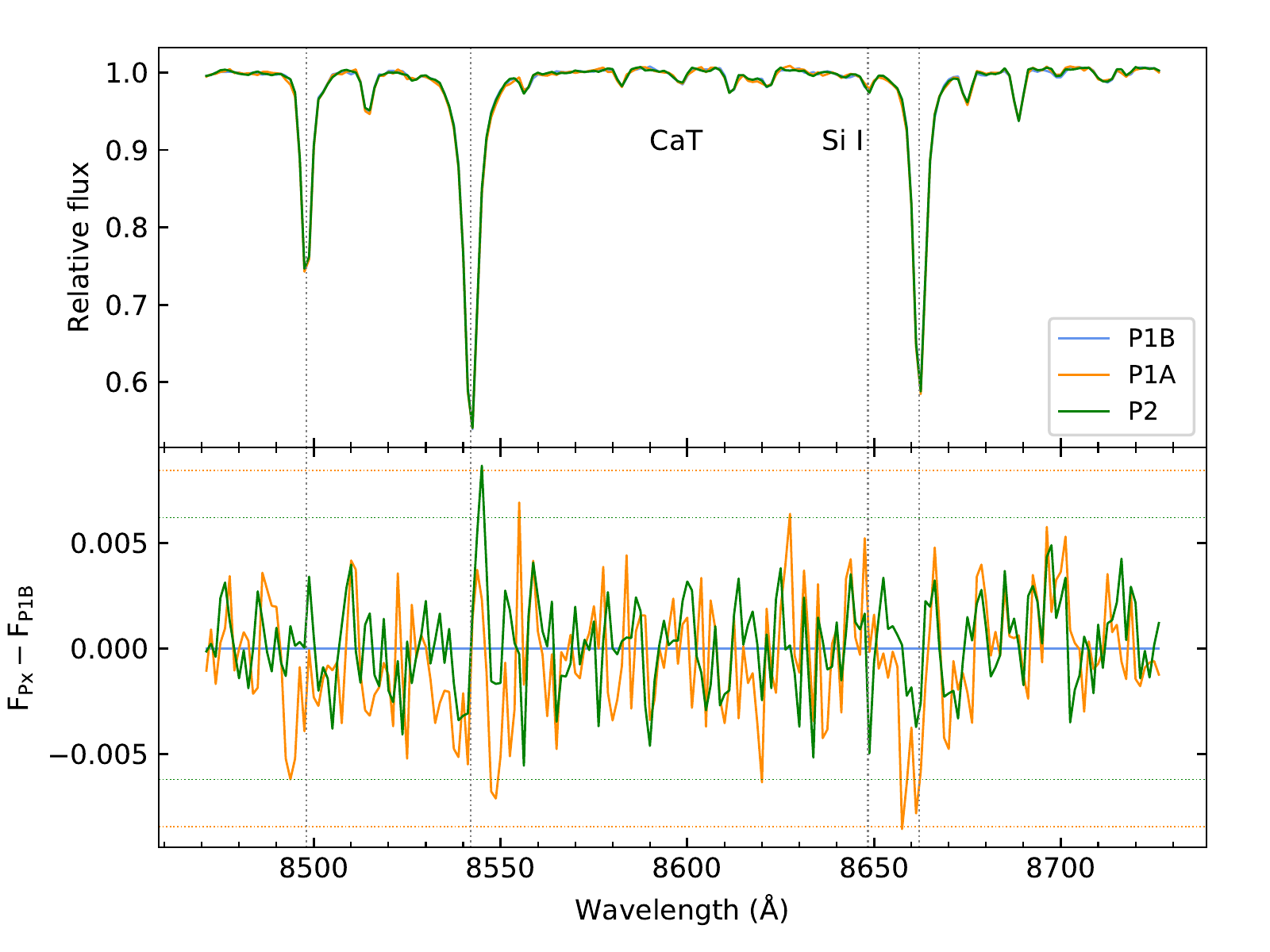}
 \includegraphics[width=\columnwidth]{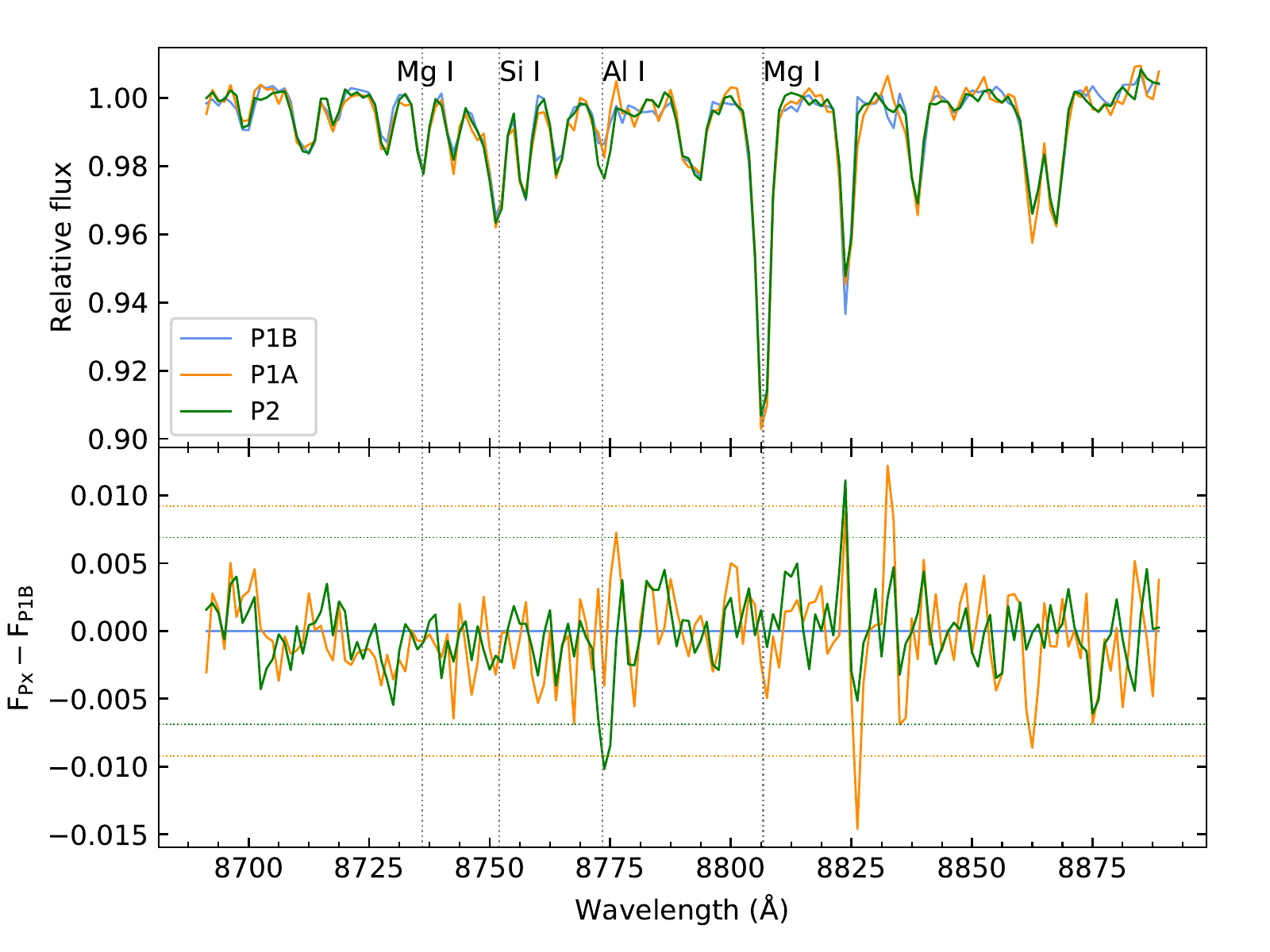}
 \caption{-- Continued.}
\end{figure*}

\end{appendix}
\end{document}